\documentclass[10pt,aps,prb,twocolumn,superscriptaddress,floatfix,noeprint,longbibliography]{revtex4-2}

\usepackage{graphicx,tabularx,makecell}
\usepackage{amsmath,amsfonts,amssymb}
\usepackage{mathtools}
\usepackage{bm,dsfont}
\usepackage[dvipsnames]{xcolor}
\usepackage[colorlinks=true, citecolor=Green, linkcolor=BrickRed, urlcolor=NavyBlue]{hyperref}
\usepackage[all]{hypcap}

\graphicspath{{figures/}}

\DeclareMathOperator{\sech}{sech}
\DeclareMathOperator{\csch}{csch}

\begin{document}

\title{Theory for Lattice Relaxation in Marginally Twisted Bilayers}

\author{Christophe De Beule}
\affiliation{Department of Physics and Astronomy, University of Pennsylvania, Philadelphia, Pennsylvania 19104, USA}
\author{Gayani N. Pallewela}
\affiliation{Centre for Advanced 2D Materials, National University of Singapore, 6 Science Drive 2, Singapore 117546}
\author{Mohammed M. Al Ezzi}
\affiliation{John A. Paulson School of Engineering and Applied Sciences, Harvard University, Cambridge, Massachusetts 02138, United States}
\author{Liangtao Peng}
\affiliation{Department of Physics, Washington University in St. Louis, St. Louis, Missouri 63130, United States}
\author{E. J. Mele}
\affiliation{Department of Physics and Astronomy, University of Pennsylvania, Philadelphia, Pennsylvania 19104, USA}
\author{Shaffique Adam}
\affiliation{Department of Physics and Astronomy, University of Pennsylvania, Philadelphia, Pennsylvania 19104, USA}
\affiliation{Department of Physics, Washington University in St. Louis, St. Louis, Missouri 63130, United States}
\affiliation{Department of Materials Science and Engineering, 
National University of Singapore, 9 Engineering Drive 1, 
Singapore 117575}

\date{\today}

\begin{abstract}
Atomically thin moir\'e materials behave like elastic membranes where at very small twist angles, the van der Waals stacking energy much exceeds the elastic energy. In this ``marginal twist" regime, the equilibrium moir\'e consists of expanded regions with low stacking energy, which cover most of the moir\'e cell, while unfavorable stackings shrink to form topological defects linked by a periodic network of domain walls. We find analytical expressions that successfully capture this strong-coupling regime for both the triangular soliton network and the honeycomb soliton network, matching predictions from \textsc{lammps} molecular dynamics simulations, and numerical solutions of continuum elasticity theory. We find an emergent universality where the theory is characterized by a single twist-angle dependent parameter. Our formalism is essential to understand experiments on a wide-range of materials of current interest including twisted bilayer graphene, both aligned and anti-aligned stacked tWSe$_2$ and tMoTe$_2$, and any other twisted homobilayer with the same stacking symmetry.
\end{abstract}

\maketitle
Moir\'e van der Waals materials are stacks of two-dimensional materials where the unit cell is engineered to be much larger than the atomic lattice constant.  This is achieved through either a lattice mismatch or a relative rotational twist between two adjacent layers.  The expansion of the unit cell acts to target and energetically flatten the lowest energy bands close to the Fermi energy. Over the past six years, such moir\'e engineering has led to a long list of exciting experimental observations including of superconductivity in twisted bilayer graphene \cite{cao_unconventional_2018} and twisted WSe$_2$ \cite{guo_superconductivity_2025,xia_superconductivity_2025}, as well as correlated states like Wigner crystals in WSe$_2$/WS$_2$ heterostructures \cite{xu_correlated_2020}, orbital Chern insulators in twisted mono-bilayer graphene \cite{polshyn_electrical_2020}, and interaction-induced halos in twisted double-bilayer graphene \cite{he_symmetry_2021}.
\begin{figure}
    \centering
\includegraphics[width=\linewidth]{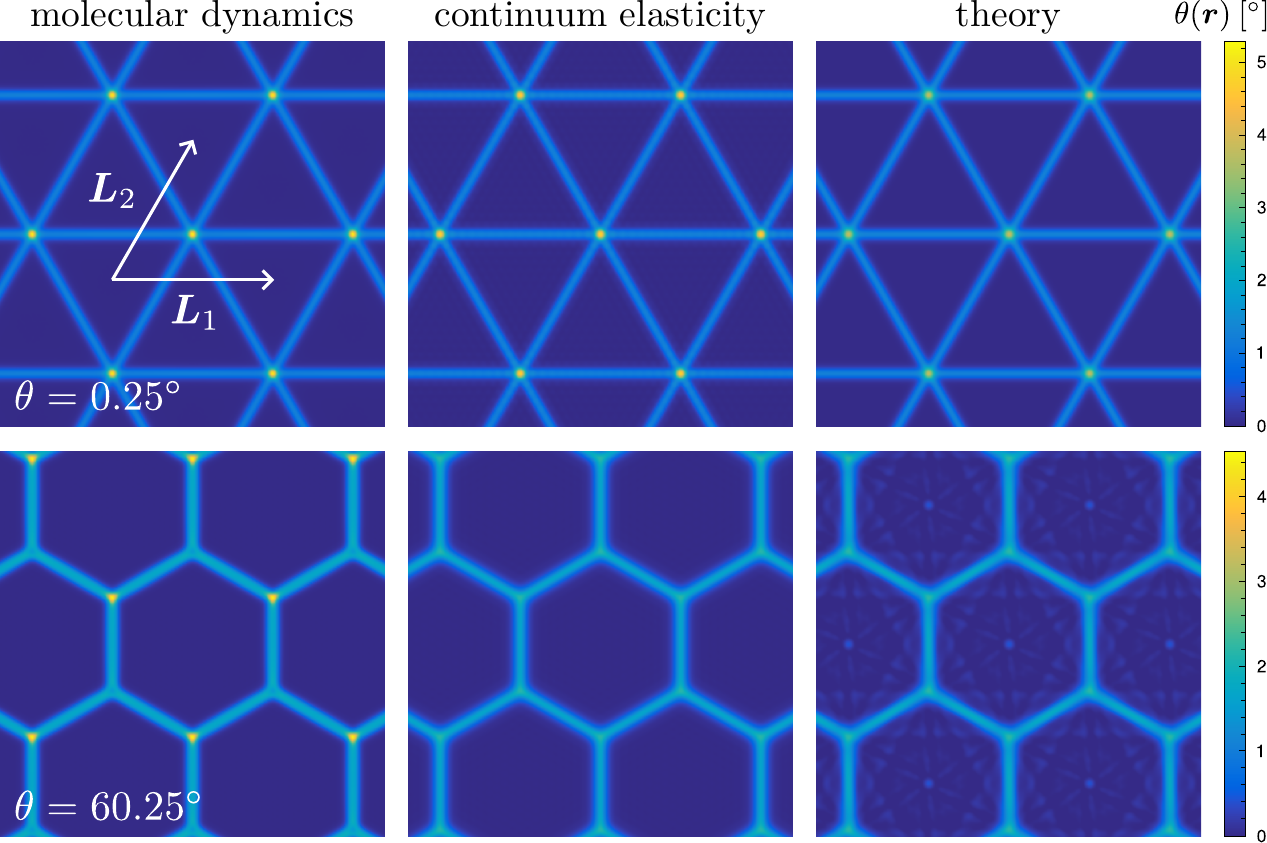}
    \caption{Comparison of our theory with \textsc{lammps} molecular dynamics simulations and numerical solutions of continuum elasticity. Top: Local twist angle $\theta(\bm r) = (1/2) \nabla \times \bm \phi(\bm r)$ for a marginal $D_6$ twist moir\'e. \textsc{lammps} simulations are for AA tWSe$_2$ with $\theta = 0.25^\circ$, continuum elasticity uses $\eta \approx 25.6$, and our theory for a triangular soliton network uses a domain wall slope $0.044$. These values are chosen to give the same physical situation. Bottom: Same for honeycomb soliton network with $\theta = 60.25^\circ$, $\eta \approx -5.9$, and domain wall slope $0.060$.}
    \label{fig:fig1}
    \vspace{-0.2in}
\end{figure}

While it is convenient to think of the constituent layers as rigid crystals, it was soon appreciated both experimentally \cite{woods_commensurateincommensurate_2014} and theoretically \cite{jung_origin_2015,san-jose_electronic_2014,van_wijk_moire_2014} that the layers behave more like flexible membranes. In this scenario, the equilibrium moir\'e structure results from ``lattice relaxation'' which arises due to the competition between intralayer elastic deformations that resist relaxation and the interlayer van der Waals (vdW) interaction that wants to expand regions with energetically favorable stacking configurations. The gain in adhesion energy grows as the moir\'e cell area leading to a significant reconstruction of the moir\'e at sufficiently small twists. Often this is an essential ingredient for the observed electronic effects. For example, without relaxation the flat bands of magic-angle twisted bilayer graphene near charge neutrality would merge with the dispersive remote bands \cite{nam_lattice_2017}, and there would not be a substrate-induced topological gap \cite{jung_origin_2015,bultinck_mechanism_2020} that underpins the observation of the orbital quantum anomalous Hall effect \cite{serlin_intrinsic_2020} in twisted bilayer graphene aligned to a hexagonal boron nitride substrate. 


To date, most treatments of structural relaxation in moir\'e materials have been numerical, relying on density-functional theory \cite{cantele_structural_2020,liu_moire_2022,noauthor_aps_nodate}, molecular dynamics simulations \cite{naik_kolmogorovcrespi_2019,thompson_lammps_2022,leconte_relaxation_2022}, or numerical solutions of continuum elasticity \cite{carr_relaxation_2018,bennett_theory_2022,nam_lattice_2017,koshino_effective_2020,kang_pseudomagnetic_2023,ramos-alonso_flat_2025}. These approaches generally work well (and are in agreement) in the  weak coupling limit, i.e.\ for ``large'' twist angles. Very recently there has also been some analytical progress in this perturbative limit \cite{ezzi_analytical_2024,ceferino_pseudomagnetic_2024,kang_analytical_2025}. In this Letter, we tackle analytically the opposite limit of strong coupling at small twist angles (the so-called ``marginal twist" regime). We find analytical expressions for the acoustic degrees of freedom that accurately match numerical simulations where the only adjustable parameters are the same as those for the large-angle theory (and thus easily obtainable numerically). As our results are asymptotically exact for both small and large twists, combining them gives a reasonably accurate theory for structural relaxation of twisted hexagonal homobilayers at all angles near $0^\circ$ or $60^\circ$.

\textcolor{NavyBlue}{\emph{Model}} --- In the long-wavelength limit, a bilayer moir\'e is defined by the local stacking or interlayer disregistry $\bm \phi(\bm r) = M \bm r + \bm u(\bm r)$ with $\bm r = (x,y)$ \cite{ochoa_moire-pattern_2019}. This is the lateral shift between the two layers, which is spatially modulated in a moir\'e material. Here $M$ is a constant $2\times2$ matrix that defines the rigid moir\'e, and $\bm u = \bm u_1 - \bm u_2$ is the acoustic displacement field, where the subscript labels the layers. Importantly, the displacement fields are defined with respect to the rigid configuration, i.e.\ two decoupled layers, and not with respect to the final equilibrium configuration. For small twists, the rigid displacement gradient $M \approx -i\theta \sigma_y + \mathcal E$ where $\mathcal E$ encodes lattice mismatch and heterostrain \cite{escudero_designing_2024}. The moir\'e lattice is defined by $\bm \phi(\bm r + \bm L) = \bm \phi(\bm r) + \bm a \equiv \bm \phi(\bm r)$ where $\bm a$ is a monolayer lattice vector, yielding $\bm L = M^{-1} \bm a$. In particular, for homobilayer twist moir\'es, we have $M \bm r = (a/L) \hat z \times \bm r$ with $L = a / [ 2 \sin (\theta / 2 ) ]$ the moir\'e lattice constant.

When the moir\'e scale much exceeds the atomic scale, we can treat the layers as continuous membranes and model the acoustic degrees of freedom with continuum elasticity \cite{nam_lattice_2017}. In this case, the static equilibrium moir\'e configuration $\bm \phi_0(\bm r)$ at zero temperature is obtained by minimizing the energy $F[\bm \phi] = F_\text{elas}[\bm \phi] + F_\text{adh}[\bm \phi]$ imposing moir\'e periodic boundary conditions. The first term gives the elastic energy cost associated with deforming two isolated layers and the second term gives the adhesion energy from the interlayer vdW interaction:
\begin{align}
    F_\text{elas}[\bm \phi] & = \int d^2 \bm r \left[ \frac{\lambda}{4} \left( \partial_i \phi_i \right)^2 + \frac{\mu}{8} \left( \partial_i \phi_j + \partial_j \phi_i \right)^2 \right], \label{eq:Felas} \\
    F_\text{adh}[\bm \phi] & = \int d^2 \bm r \, V[\bm \phi(\bm r)],
\end{align}
where $\mu, \lambda$ are monolayer Lam\'e constants, $V(\bm \phi)$ is the stacking-fault energy \cite{carr_relaxation_2018} or adhesion potential, and summation is implied. Here we have assumed that quadratic out-of-plane contributions to the strain tensor, defined relative to isolated layers, is negligible compared to the linear in-plane part. In this case, the layer center-of-mass [$(\bm u_1 + \bm u_2)/2$] and relative ($\bm u_1 - \bm u_2$) motion are decoupled. The latter gives Eq.\ \eqref{eq:Felas} up to constants and boundary terms \footnote{For a twist moir\'e, the rigid displacement gradient is antisymmetric: $M_{ij} = -M_{ji}$ and $\phi_{ij} = (\partial_i \phi_j + \partial_j \phi_i)/2 = u_{ij}$}, see Supplemental Material (SM) \footnote{See Supplemental Material at [insert link] for more details on the methodology employed to solve the equations of motion from continuum elasticity, which includes Refs.\ \cite{mesple_giant_2023,pulay_improved_1982,jiang_parametrization_2015}.}.
Note also that this theory only describes acoustic displacements which are dominant in the regime we consider. Corrections from internal relaxation in the monolayer cell (optical displacements) are of order $a^2/L$ \cite{de_beule_elastic_2025}.

For twist moir\'es in particular, the theory is characterized by a coupling constant $\eta = (L^2/a^2) V_1 / \mu \approx c_1/\theta^2$ with $c_1 = V_1/\mu$. Here $V_1$ is the dominant Fourier component of the adhesion potential, which sets the energy barrier between favorable and unfavorable stackings. One then identifies two regimes: a weak coupling ``large angle'' regime $|\eta| \ll 1$ and a strong coupling ``small angle'' regime $|\eta| \gg 1$.  For weak coupling, we have shown in a recent Letter \cite{ezzi_analytical_2024} that the displacement field for twist moir\'es with $D_6$ or $D_3$ symmetry in lowest order is
\begin{equation} \label{eq:first}
    \bm u(\bm r) = \frac{\sqrt{3}a|c_1|}{\pi \theta^2} \sum_{i=1}^3 \hat z \times \hat g_i \sin(\bm g_i \cdot \bm r + \psi_1),
\end{equation}
where $\psi_1 = \arg(c_1)$ with $\psi_1 = 0,\pi$ for $D_6$. Here layer 1 and 2 are rotated counterclockwise by $\pm \theta/2$, respectively, and $\bm g_i$ ($i=1,2,3$) are moir\'e reciprocal vectors of the first star related by $120^\circ$ in-plane rotations. This theory can be refined by going to higher order in $\eta$ \cite{ezzi_analytical_2024,kang_analytical_2025}.

In this work, we instead consider the strong coupling regime and develop an expansion of the displacement field in $1/\eta = \theta^2/c_1$. In this limit, it is well known that domains with near uniform stacking emerge on the scale of the moir\'e lattice \cite{alden_strain_2013,tilak_moire_2023}, see Fig.\ \ref{fig:fig1}. These correspond to energetically favorable stacking configurations, while unfavorable ones contract, forming topological defects linked by a periodic network of domain walls (solitons in $\bm \phi$). In this regime, the moir\'e period is much larger than the domain wall width, and $\sqrt{|c_1|}/\theta$ is exactly the ratio of these lengths \cite{ramos-alonso_flat_2025}. Examples of twist moir\'es with $D_6$ stacking are twisted bilayer graphene ($|c_1| \sim 10^{-5}$) and homobilayers with $D_{3h}$ ($D_{3d}$) layers twisted near $0^\circ$ ($60^\circ$). These include twisted double bilayer graphene with ABBA stacking \cite{koshino_band_2019} as well as 2H (1T) transition metal dichalcogenides (TMDs) ($|c_1| \sim 10^{-4}$). While twisting these near $60^\circ$ ($0^\circ$) yields structures with $D_3$ stacking. We note that $c_{1,\text{tTMD}} / c_{1,\text{tBG}} \sim 10$ such that the similar moir\'e reconstruction is expected for $\theta_\text{tTMD} \approx 3 \theta_\text{tBG}$.

To gain insight into the strong coupling limit, we consider a first-star theory with $D_6$ symmetry. Up to an additive constant, $V[\bm \phi] = 2 V_1 \sum_{i=1}^3 \cos(\bm g_i M^{-1} \cdot \bm \phi)$. When $V_1 > 0$, the potential has two degenerate minima denoted as AB and BA stacking ($\bm \phi = \pm a \hat x/\sqrt{3}$) and one maxima which we call AA stacking ($\bm \phi = \bm 0$). Note that AA need not correspond physically to eclipsing layers. 
Because these extrema carry index $1$, the Poincar\'e-Hopf theorem requires the existence of three saddle points (SP) with index $-1$. This results in a {\it triangular network} of solitons with domain walls at the SPs. By contrast, when $V_1 < 0$, AA stacking is the only favorable configuration, and one obtains a {\it honeycomb} network. The latter also applies to $D_3$ structures for $|\eta| \gg 1$. While there are two unfavorable nondegenerate high-symmetry configurations for $D_3$ stacking symmetry, both contract to a point resulting in an emergent $D_6$ symmetry. 
\begin{figure}
    \centering
    \includegraphics[width=\linewidth]{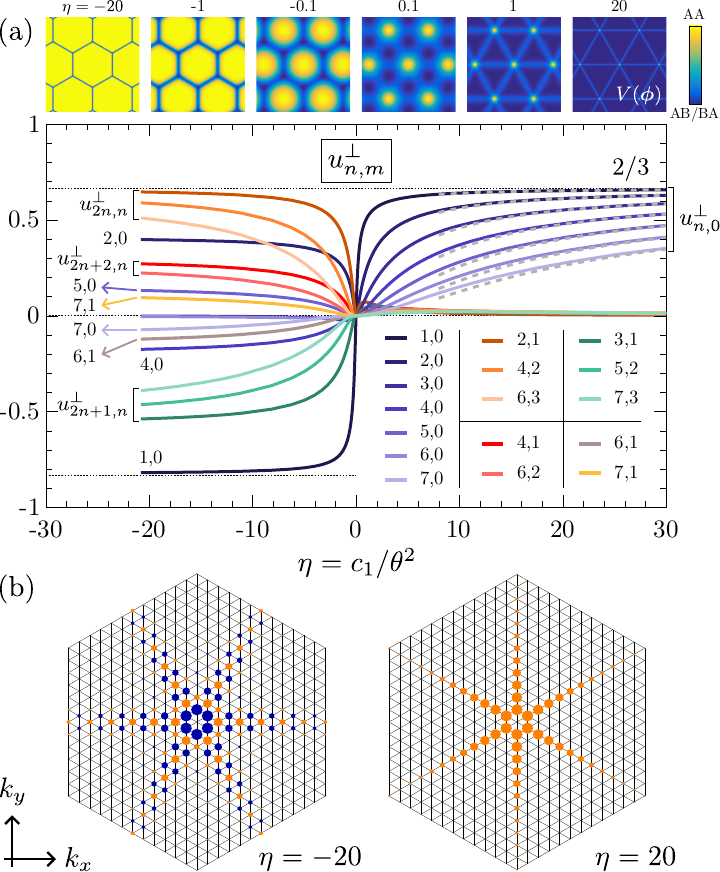}
    \caption{(a) Transverse components $u^\perp_{n,m}$ as a function of $\eta = c_1/\theta^2$ where $m=0,\ldots,n-1$ for stars with representatives $n \bm g_1 + m \bm g_2$. We only show distinct and appreciable values for the first 7 shells. Solid (dashed) lines give results from continuum elasticity (theory). The top panel shows $V[\bm \phi(\bm r)]/V_1$ in real space. (b) Reciprocal structure $u^\perp_{\bm g}$ up to $12$ shells ($78$ stars) for $\eta = \mp 20$ where the dot size and color give the magnitude and sign (orange positive and blue negative).}
    \label{fig:fig2}
\end{figure}

It is useful to define longitudinal and transverse Fourier components of the displacement field $u_{\bm g}^\parallel = (iL/a) \bm g \cdot \bm u_{\bm g}$ and $u_{\bm g}^\perp = (iL/a) ( \hat z \times \bm g ) \cdot \bm u_{\bm g}$. These give the divergence and curl, respectively. 
Under a moir\'e stacking symmetry $\mathcal S$ we then find $u_{\mathcal S \bm g}^\parallel = \pm u_{\bm g}^\parallel$ and $u_{\mathcal S \bm g}^\perp = \pm \det(\mathcal S) u_{\bm g}^\perp$ with an extra sign when the layers are swapped. 
Moreover, they decouple in the elastic energy density:
\begin{equation}
    f_\text{elas}(u_{\bm g}^\parallel,u_{\bm g}^\perp) = \frac{a^2}{4L^2} \sum_{\bm g} \left[ \left( \lambda + 2 \mu \right) | u_{\bm g}^\parallel |^2 + \mu | u_{\bm g}^\perp |^2 \right],
\end{equation}
up to constants which vanish for a pure twist.

Because twist moir\'es relax mostly by local twisting (twirling) \cite{kazmierczak_strain_2021,ezzi_analytical_2024} the $u^\perp_{\bm g}$ are dominant. We thus set $\lambda = 0$ as it does not change the essential physics we are interested in. Numerically, we find $|u_{\bm g}^\parallel/u_{\bm g}^\perp| < 10^{-2}$ for $\lambda \geq 0$. For a $D_6$ twist moir\'e, $\mathcal C_{6z}$ yields one real $u^\parallel_{n,m}$ and $u^\perp_{n,m}$ for each star with representative $n\bm g_1 + m \bm g_2$ belonging to the $n$th shell ($m=0,\ldots,n-1$). Here we define stars as equal length reciprocals closed under $\mathcal C_{6z}$ with $\bm g_1 = 4\pi \hat y/3L$ and $\bm g_2 = \mathcal C_{3z}\bm g_1$. In addition, $\mathcal C_{2x}$ gives $u^\parallel_{n,m} = -u_{n,n-m}^\parallel$ and $u^\perp_{n,m} = u_{n,n-m}^\perp$. Hence longitudinal components vanish for $\mathcal C_{2x}$ invariant stars.

\textcolor{NavyBlue}{\emph{Triangular soliton network}} --- For the triangular network, the shear solitons are partial dislocations (AB/SP/BA) and the \emph{exact} solution is
\begin{equation} \label{eq:utri}
    \bm u(\bm r) = \frac{\sqrt{3}a}{2\pi} \sum_{i=1}^3 \hat z \times \hat g_i \sum_{n=1}^\infty \frac{u_{n,0}^\perp}{n} \sin(n \bm g_i \cdot \bm r),
\end{equation}
where $\bm g_3 = -\bm g_1 - \bm g_2$.
This field has support on stars concentric with the first that are all orthogonal to the domain walls and by symmetry $u_{n,0}^\parallel = 0$.  Consider first the limit $\eta \rightarrow \infty$. Although continuum elasticity breaks down, it is instructive. We find $u_{n,0}^\perp \rightarrow 2/3$ and
\begin{equation} \label{eq:utrilimit}
    \lim_{\eta\rightarrow\infty} \bm u(\bm r) = \frac{a}{\sqrt{3}\pi} \sum_{i=1}^3 \hat z \times \hat g_i \arctan\left[ \cot \left( \frac{ \bm g_i \cdot \bm r}{2} \right) \right],
\end{equation}
giving sharp domain walls. The factor $2/3$ is determined by the shift between AA and AB/BA stacking, which is the largest possible displacement. For finite $\eta \gg 1$, the domain walls attain a finite width. We match their slope to that of a single AB/SP/BA domain wall centered at $\bm r_0 = \sqrt{3} L \hat y / 2$ \cite{gao_symmetry_2022}. Expanding Eq.\ \eqref{eq:utri} in $y/L$ gives
\begin{equation}
    \bm \phi(\bm r_0 + y \hat y) - \bm \phi_\text{SP} \simeq -\frac{y \hat xa}{L} \left( 1 + \sum_{n=1}^\infty \left[ 2 + (-1)^n \right] u^\perp_{n,0} \right),
\end{equation}
which needs to match the slope $2\sqrt{\eta} a/L = 2 \sqrt{c_1}$ of the isolated soliton, as shown in the SM. This is satisfied by
\begin{equation} \label{eq:uperp}
    u^\perp_{n,0}(\theta) = \frac{2}{3} \sech^2 \left( \frac{2n \theta}{3 \sqrt{c_1}} \right),
\end{equation}
among other choices \cite{Note2}. This function decays exponentially with $n$ and the sum converges for $n_\text{max} \approx \eta$.
\begin{figure}
    \centering
    \includegraphics[width=\linewidth]{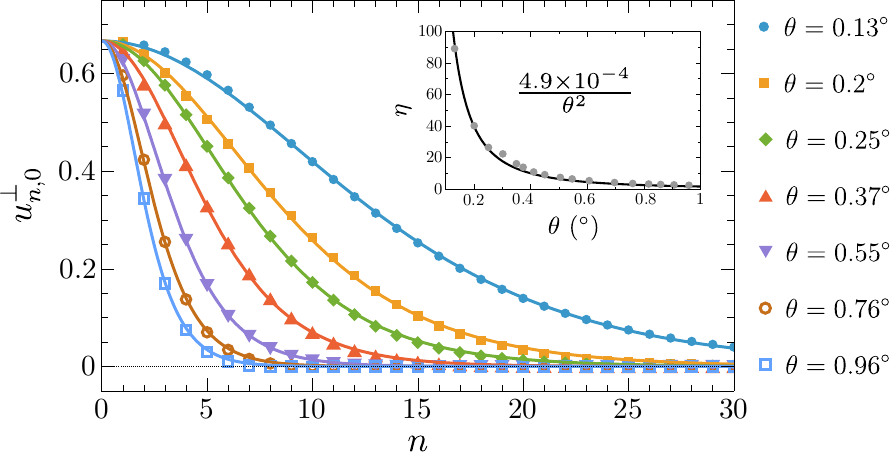}
    \caption{Dominant Fourier components $u_{n,0}^\perp$ obtained from \textsc{lammps} simulations for aligned tWSe$_2$ and fit to $(2/3) \sech^2 \left[ 2n/(3\sqrt{\eta}) \right]$. The inset gives the fitted values of $\eta$ as a function of twist angle, showing the expected $1/\theta^2$ behavior with a domain wall slope $2\sqrt{c_{1,\text{eff}}} \approx 0.044$.}
    \label{fig:fig3}
\end{figure}

In Fig.\ \ref{fig:fig1} we compare Eq.\ \eqref{eq:utri} to both the numerical solution of continuum elasticity and to \textsc{lammps} molecular dynamics simulations with state-of-the-art interatomic potentials~\cite{Note2}. Including second and third star contributions to the adhesion potential modifies the domain wall slope to $2\sqrt{c_1-8c_2+9c_3}$ with $c_i = V_i/\mu$ \cite{ramos-alonso_flat_2025,Note2}.  We can capture these contributions within the first-star theory using an effective $c_{1,\text{eff}} = c_1 -8c_2 +9c_3$, e.g., for tWSe$_2$ near 0$^\circ$ we find $c_{1,\text{eff}} \approx 4.88 \times 10^{-4}$ matching approximately the value obtained by extracting the $c_i$ in the large-angle regime \cite{Note2}. While our theory accurately reproduces the domain wall, the size of the AA region is slightly overestimated because of small contributions from other stars and because Eq.\ \eqref{eq:uperp} slightly deviates from the (exponentially decaying) numerical values at larger $n$. Nonetheless we find remarkably good agreement with the numerical solution [see dashed lines in Fig.\ \ref{fig:fig2}(a)]. Moreover, in Fig.\ \ref{fig:fig3} we show that our \textit{ansatz} for the transverse components fits perfectly \textsc{lammps} simulations for aligned tWSe$_2$ where the only fit parameter is the slope of the domain wall.
\begin{figure}
    \centering
    \includegraphics[width=\linewidth]{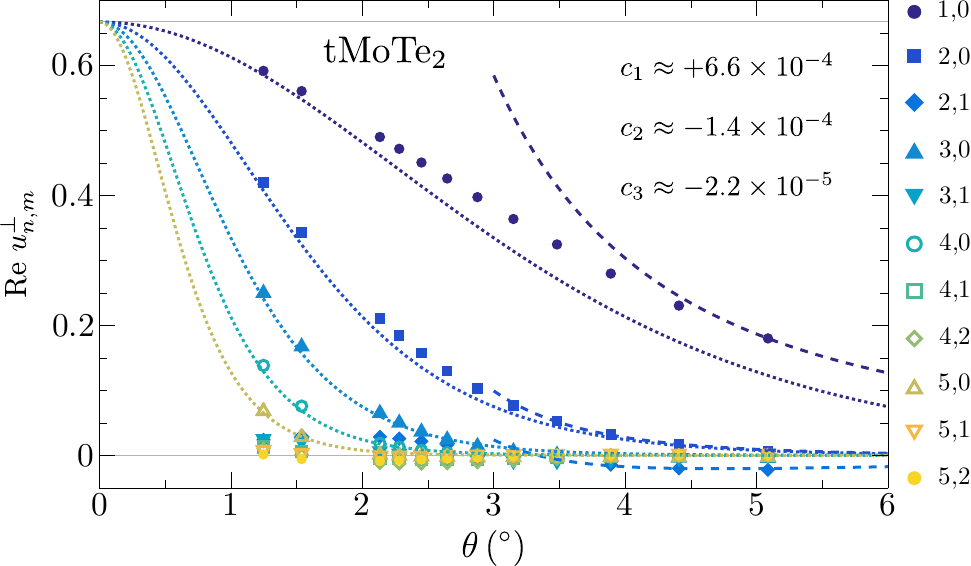}
    \caption{Real part of transverse Fourier components $u_{n,m}^\perp$ of the acoustic in-plane displacement field for tMoTe$_2$. Data points were obtained from Ref.\ \cite{zhang_polarization-driven_2024} where machine learning force fields were used together with large-scale density-functional theory calculations. The dashed lines are fits to 2nd order perturbation theory in $c_i/\theta^2$  \cite{Note2} giving $c_1$, $c_2$, and $c_3$ as indicated, and dotted lines show the small-angle theory $u^\perp_{n,0} = (2/3) \sech^2 \left[ 2n \theta /(3\sqrt{c_{1,\text{eff}}}) \right]$
    with {\it no further adjustable parameters}, i.e., $c_{1,\text{eff}} = c_1 - 8c_2 + 9c_3 \approx 1.58 \times 10^{-3}$.}
    \label{fig:fig4}
\end{figure}

\textcolor{NavyBlue}{\emph{Honeycomb soliton network}} --- In the honeycomb case the solitons are full shear dislocations (AA/SP/AA). Because now the network is not primitive, the solution is much more complicated. Unlike for the triangular case, for which there is only one type of dominant contribution, we find (see Fig.\ \ref{fig:fig2}) that the main contributions for the honeycomb network are: (i)  $u_{n,0}^\perp$ which oscillate and decay rapidly with $n$, and (ii) three other sets of transverse stars: $u^\perp_{2n,n}$, $u^\perp_{2n+1,n} = u^\perp_{2n+1,n+1}$, and $u^\perp_{2n+2,n} = u^\perp_{2n+2,n+2}$. The set $u^\perp_{2n,n}$ by itself yields a triangular network scaled by $1/\sqrt{3}$ and rotated by $30^\circ$, giving too many solitons with the wrong slope. The role of the remaining contributions is to remove the extra ones and correct the slope.

As before we can expand $\bm \phi(\bm r)$ near a domain wall $\bm r_0 = \hat xL/2$, retaining only these stars. Setting the slope equal to that of the isolated soliton gives
\begin{equation}
    \begin{aligned}
        & 4 \sqrt{2|\eta|} \approx 1 + 3\sum_{n=1}^\infty (-1)^n u_{n,0}^\perp + \frac{2 + (-1)^n}{3} \, u_{2n,n}^\perp \\
        & \quad - \frac{(1 + 2n)[1 + (-1)^n + 2n]}{1+3n(1+n)} \, u_{2n+1,n}^\perp \\
        & \quad + \frac{2(1+n)^2 + (-1)^n (2+2n+n^2)}{2+3n(1+n/2)} \, u_{2n+2,n}^\perp,
    \end{aligned}
\end{equation}
where $4\sqrt{2|c_1|}$ is the slope of an isolated AA/SP/AA soliton \cite{Note2}. However, because there are multiple contributions, this gives too few conditions. Therefore, we take a different approach. First, we extract the domain wall slope from the \textsc{lammps} simulations which gives an effective $\eta$. For tWSe$_2$ with $\theta = 60.25^\circ$ we find $\eta \approx -5.9$ corresponding to $c_{1,\text{eff}} \approx -1.14 \times 10^{-4}$. We then fit continuum elasticity according to
\begin{align}
    u_{n,0}^\perp & = -0.26 \sin \left( 2.1 n \right) \csch(0.27 n), \\
    u_{2n+i,n}^\perp & = \left( a_i + \frac{b_i}{\sqrt{|\eta|}} \right) \sech^2 \left[ n \left( r_i + \frac{s_i}{\sqrt{|\eta|}} \right) \right],
\end{align}
where the parameters are given in Table \ref{tab:params}.
We contrast this to the triangular case where using the same procedure gives $(a,b,r,s)=(0.66,0,0.021,0.55)$.  We note that Eq.~\ref{eq:uperp} corresponds to $(a,b,r,s)=(2/3,0,0,2/3)$.  While we believe that the values in Table \ref{tab:params} are similarly related, we have not been able to find such a simplification.
\begin{table}
    \label{tab:params}
    \centering
    \begin{tabular}{c | c | c | c | c}
        \Xhline{1pt}
        & $a$ & $b$ & $r$ & $s$ \\
        \hline
        $u^\perp_{2n,n}$ & $0.68$ & $-0.096$ & $0.028$ & $0.64$ \\ 
        $u^\perp_{2n+1,n}$ & $-0.59$ & $0.23$ & $0.033$ & $0.69$ \\ 
        $u^\perp_{2n+2,n}$ & $0.31$ & $-0.20$ & $0.042$ & $0.69$ \\ 
        \Xhline{1pt}
    \end{tabular} \qquad
    \caption{Parameters for honeycomb soliton network obtained by fitting to continuum elasticity for $\eta \in [-25,-10]$.} 
\end{table}

\textcolor{NavyBlue}{\emph{Discussion}} --- To illustrate the power of the analytical theory, we apply it to aligned tMoTe$_2$. This is an important example since this material hosts exotic phases like fractional quantum anomalous Hall states \cite{cai_signatures_2023,zhang_polarization-driven_2024}. Here relaxation is particularly important since it changes the band topology. The experimentally observed Chern number sequence \cite{kang_evidence_2024} was reproduced by the machine learning force field method of Ref.\ \cite{zhang_polarization-driven_2024}.
To test our theory, we first extract the dominant acoustic part of the in-plane displacement field from Ref.\ \cite{zhang_polarization-driven_2024}. The result is shown in Fig.\ \ref{fig:fig4} where we only show the real part of the transverse Fourier components because the imaginary and longitudinal parts are three orders of magnitude smaller. Small imaginary parts are present because $D_6$ is only an approximate symmetry for 2H tTMDs twisted near $0^\circ$.
Consistent with Fig.\ \ref{fig:fig2} the small-angle regime is dominated by $u_{n,0}^\perp$ which signals domain formation. We then obtain the parameters $c_i = V_i / \mu$ by fitting the data for large angles to perturbative solutions of continuum elasticity up to 2nd order in $c_i/\theta^2$ \cite{Note2}. 
Without further fitting, we compare the data at small twist angles to the analytical theory of Eq.\ \eqref{eq:uperp} using $c_{1,\text{eff}} = c_1-8c_2+9c_3$. As shown in Fig.\ \ref{fig:fig4}, the agreement is remarkable. We emphasize that the effective parameter for small twists is completely determined by the fit at large twist angles. 

Our theoretical framework provides analytic expressions for the in-plane center of mass positions, at both large \cite{ezzi_analytical_2024} and small (present work) twist angles for a moir\'e homo-interface built from layers with $D_{3d}$ and $D_{3h}$ symmetry. Thus, lattice relaxation can be systematically included into electronic continuum models \cite{foutty_mapping_2024} previously developed for rigid lattices at high symmetry points, e.g., Refs.\ \cite{lopesdossantos_graphene_2007,bistritzer_moire_2011} for $K/K'$, and Refs.\ \cite{lei_moire_2024,calugaru_new_2024} at the $M$ points. In principle, this will yield accurate fully-relaxed electronic structures of moir\'e materials at any twist angle including in the marginal-twist regime. 

The solutions presented here are valid for any marginally stacked 2D material where the stacking energy landscape has either $D_3$ or $D_6$ symmetry. As discussed earlier, this already encompasses a wide range of existing (and yet to be discovered) twisted 2D materials including homobilayers of graphene and TMDs in both aligned and antialigned configurations. Moreover, our procedure can also be applied to systems with square and rectangular symmetry, e.g.\ PdSe$_2$ \cite{zhang_quantum_2024} or GeSe \cite{kennes_one-dimensional_2020}. In general, twisted homobilayers with $D_{4d}$ or $D_{4h}$ layers are dual to hexagonal ones with $D_{3h}$ or $D_{3d}$ symmetry, respectively.

This highlights an important aspect of the strong coupling limit -- it is governed by a new emergent universality, i.e., models with very different properties and symmetries at large twist angle can behave very similarly at small twist angle.  The physical properties depend only on the symmetry of the soliton domain wall network and a single (twist-angle dependent) effective parameter given by the ratio of the moir\'e length to the domain wall width.  Hints of this universality is already present in existing experiments.  As seen in Fig.\ \ref{fig:fig4}, for tTMDs, the strong coupling physics already becomes important for $\theta \lesssim 3^\circ$. Therefore, it is unsurprising that in this regime DFT predictions deviate strongly from continuum models extrapolated from large twist angles \cite{devakul_magic_2021, zhang_polarization-driven_2024}. Moreover, this helps explain the similarity between the quantum spin Hall states observed in both twisted WSe$_2$ at 3$^\circ$ and MoTe$_2$ at 2.1$^\circ$ \cite{kang_evidence_2024,kang_double_2024}. Since these are both in the strong coupling regime for lattice relaxation, we expect that the electronic properties are universal and governed mostly by the emergent triangular soliton network.

\let\oldaddcontentsline\addcontentsline 
\renewcommand{\addcontentsline}[3]{} 
\begin{acknowledgments}
We thank Valentin Cr\'epel and Di Xiao for interesting discussions and Martin Claassen for computational resources. The \textsc{lammps} simulations were performed at the National University of Singapore High Performance Computing facility (NUSREC-HPC-00001). We also thank the authors of Ref.\ \cite{zhang_polarization-driven_2024} for providing us with their full numerical data set using machine learning force fields. The \textsc{Julia} code used to solve the continuum elasticity equations of motion self-consistently is available on a public GitHub repository \cite{debeule2026}. LAMMPS data are available upon reasonable request. L.P.\ and S.A.\ are supported by a start-up grant at Washington University in St.\ Louis. G.N.P., L.P., and S.A.\ acknowledge support of Singapore National Research Foundation Investigator Award (NRF-NRFI06-2020-0003). M.M.A.E acknowledges support from the Ministry of Education, Singapore (Research Centre of Excellence award to the Institute for Functional Intelligent Materials, I-FIM, project No. EDUNC-33-18-279-V12), the National Research Foundation, Singapore, under its AI Singapore Programme (AISG Award No: AISG3-RP-2022-028) and Simons Foundation Award No.\ 896626. C.D.B.\ and E.J.M.\ are supported by the U.S.\ Department of Energy under Grant No.\ DE-FG02-84ER45118. 
\end{acknowledgments}

\bibliography{references}
\let\addcontentsline\oldaddcontentsline 


\clearpage
\onecolumngrid
\begin{center}
\textbf{\Large Supplemental Material for ``Elastic Screening of Pseudogauge Fields in Graphene''}
\end{center}
 
\setcounter{equation}{0}
\setcounter{figure}{0}
\setcounter{table}{0}
\setcounter{page}{1}
\setcounter{secnumdepth}{2}
\makeatletter
\renewcommand{\thepage}{S\arabic{page}}
\renewcommand{\thesection}{S\arabic{section}}
\renewcommand{\theequation}{S\arabic{equation}}
\renewcommand{\thefigure}{S\arabic{figure}}
\renewcommand{\thetable}{S\arabic{table}}

\tableofcontents
\vspace{1cm}

\twocolumngrid

\section{Continuum elasticity} \label{sm:elasticity}

We start by deriving the equations of motion governing the disregistry $\bm \phi(\bm r)$ in the absence of out-of-plane corrugations. The equilibrium configuration is obtained from the variational derivative
\begin{equation}
    \frac{\delta F}{\delta \bm \phi(\bm r)} = 0, \qquad F = \int d^2\bm r \, \mathcal F\left[ \bm \phi(\bm r), \nabla \bm \phi(\bm r) \right],
\end{equation}
which gives
\begin{equation}
    \frac{\partial \mathcal F}{\partial \phi_i} - \partial_j \frac{\partial \mathcal F}{\partial ( \partial_j \phi_i )} = 0.
\end{equation}
One finds
\begin{equation}
    \frac{\partial \mathcal F}{\partial ( \partial_j \phi_i )} = \frac{\lambda}{2} \delta_{ij} \partial_i \phi_i + \frac{\mu}{2} \left( \partial_j \phi_i + \partial_i \phi_j \right),
\end{equation}
where repeated indices are \textit{not} summed. This yields  \cite{gao_symmetry_2022,ramos-alonso_flat_2025}
\begin{equation}
    2 \frac{\partial V}{\partial \bm \phi} = \left( \lambda + \mu \right) \nabla ( \nabla \cdot \bm \phi ) + \mu \nabla^2 \bm \phi,
\end{equation}
with $V[\bm \phi(\bm r)]$ the stacking-fault energy, $\mu$ the shear modulus, and $\mu + \lambda$ the bulk modulus. However, since the vector Laplacian satisfies
\begin{equation}
    \nabla^2 \bm \phi = \nabla \left( \nabla \cdot \bm \phi \right) - \nabla \times \left( \nabla \times \bm \phi \right),
\end{equation}
we can write the equations of motion as
\begin{equation}
    \mu \nabla \times \left( \nabla \times \bm \phi \right) - \left( \lambda + 2 \mu \right) \nabla \left( \nabla \cdot \bm \phi \right) = -2 \frac{\partial V}{\partial \bm \phi}.
\end{equation}
This separates the rotational and volumetric components on the left-hand side.
\begin{figure}
    \centering
    \includegraphics[width=\linewidth]{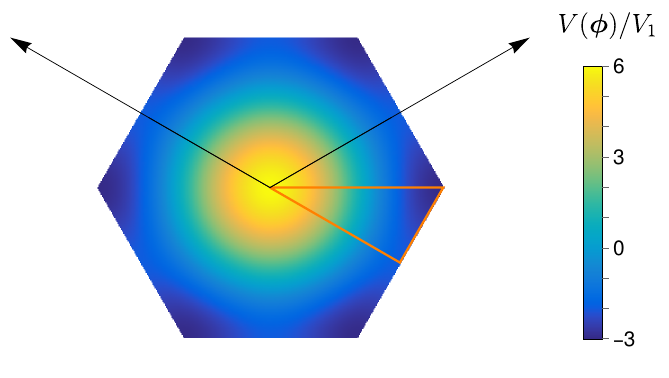}
    \caption{Adhesion potential with $C_{6v}$ symmetry in the first-star approximation versus the stacking configuration $\bm \phi = (\phi_x, \phi_y)$. Arrows are monolayer primitive lattice vectors and the orange triangle contains all independent configurations.}
    \label{fig:sfeD6}
\end{figure}

In the first-star approximation, the stacking-fault energy with $C_{6v}$ symmetry is given by
\begin{equation}
    V(\bm \phi) = V_0 + 2 V_1 \sum_{i=1}^3 \cos \left( \bm b_i \cdot \bm \phi \right),
\end{equation}
with $V_0, V_1$ real constants and where $\bm b_i$ are the shortest nonzero reciprocal vectors of the monolayer that are related by threefold rotations. Since only derivatives of $V(\bm \phi)$ appear in the equations of motion, we set $V_0 = 0$ from now on (as we do in the main text). The stacking-fault energy is shown in Fig.\ \ref{fig:sfeD6}. We proceed to solve the equations of motion for a single isolated domain wall for both cases: $V_1 > 0$ and $V_1 < 0$. In particular, we consider two semi-infinite domains with translational symmetry along the direction of the domain wall. We then consider the full problem for a moir\'e bilayer where we solve the equations of motion numerically in Fourier space using the self-consistent method implemented in \textsc{julia} in conjunction with the \textsc{diis} method (direct inversion of the iterative subspace).

\subsection{AB/BA domain walls}

We first consider the case $V_1 > 0$. We take a domain wall located at $y = 0$ with translational symmetry along $x$ that separates regions with AB ($y > 0$) and BA ($y < 0$) stacking. Hence $\bm \phi = \bm \phi(y)$ and the equations of motion become
\begin{equation} \label{eq:eomwall}
    \begin{bmatrix} \mu \phi_x'' \\ \left( \lambda + 2 \mu \right) \phi_y'' \end{bmatrix} = -4 V_1 \sum_{i=1}^3 \bm b_i \sin \left( \bm b_i \cdot \bm \phi \right),
\end{equation}
where the primes indicate derivatives with respect to the $y$ coordinate. In the coordinate system of Fig.\ 1 of the main text, we can take $\bm b_1 = 4\pi \hat x / \sqrt{3}a$ and $\bm b_{2/3}$ related by $\mathcal C_{3z}$. Setting $a=1$ we explicitly have
\begin{widetext}
\begin{equation}
    \sum_{i=1}^3 \bm b_i \sin \left( \bm b_i \cdot \bm \phi \right) = \frac{4\pi}{\sqrt{3}} \begin{pmatrix} \sin \left( \frac{2\pi \phi_x}{\sqrt{3}} \right) \left[ \cos \left( 2\pi \phi_y \right) + 2 \cos \left( \frac{2\pi \phi_x}{\sqrt{3}} \right) \right] \\ \sin\left( 2 \pi \phi_y \right) \cos \left( \frac{2\pi \phi_x}{\sqrt{3}} \right) \end{pmatrix}.
\end{equation}
\end{widetext}

\begin{figure}
    \centering
    \includegraphics[width=\linewidth]{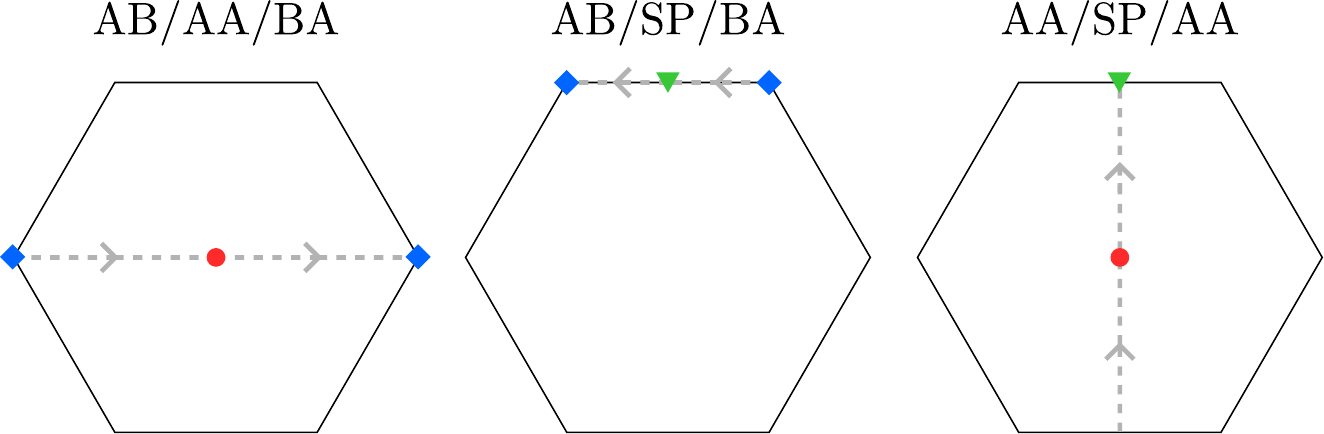}
    \caption{Three domain walls shown in configuration space (the $\phi_x-\phi_y$ plane) as dashed lines where the arrows point from $-\infty$ to $+\infty$ in the direction perpendicular to the domain wall. The red dot, blue diamonds, and green triangle correspond to AA, AB/BA, and SP stacking, respectively.}
    \label{fig:dw_config}
\end{figure}

\begin{figure}
    \centering
    \includegraphics[width=\linewidth]{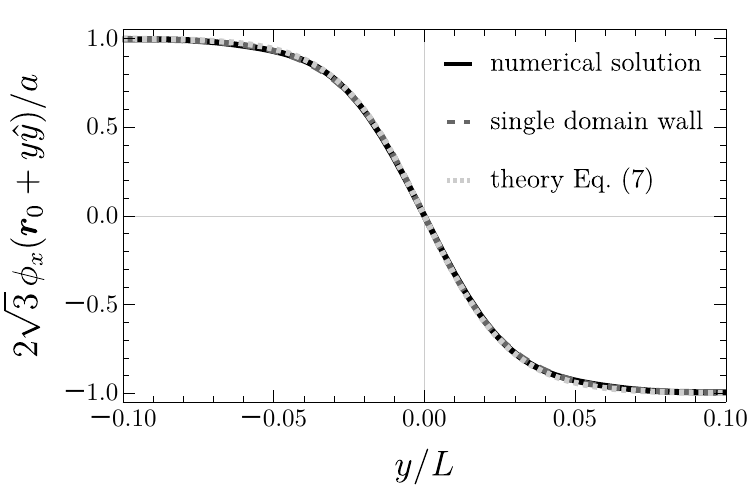}
    \caption{AB/SP/BA domain wall of the triangular network ($V_1>0$) comparing the numerical solution for the full moir\'e from continuum elasticity to the analytical result for a single domain wall, and the theory in Eq.\ (5) of the main text.}
    \label{fig:dw}
\end{figure}

We now consider two different types of pure shear domain walls, which are illustrated in Fig.\ \ref{fig:dw_config}. The first type is an AB/AA/BA interface with $\phi_y = 0$ and boundary conditions
\begin{equation}
    \bm \phi(y = \pm \infty) = \frac{1}{\sqrt{3}} \begin{pmatrix} \pm 1 \\ 0 \end{pmatrix} = \bm \phi_{\text{AB/BA}},
\end{equation}
and the second one is an AB/SP/BA interface with $\phi_y = 1/2$ and boundary conditions
\begin{equation}
    \bm \phi(y = \pm \infty) = \frac{1}{\sqrt{3}} \begin{pmatrix} \mp \frac{1}{2} \\ \frac{\sqrt{3}}{2} \end{pmatrix} = \bm \phi_{\text{AB/BA}},
\end{equation}
For both types of domain wall, the constant $\phi_y$ solves one of the equations in Eq.\ \eqref{eq:eomwall}.
We are left with a second-order ordinary differential equation
\begin{equation}
    \phi_x'' + \frac{16 c_1 \pi}{\sqrt{3}} \left[ \pm \sin \left( \frac{2\pi \phi_x}{\sqrt{3}} \right) + \sin \left( \frac{4\pi \phi_x}{\sqrt{3}} \right) \right] = 0,
\end{equation}
with $c_1 = V_1 / \mu$ and where $\pm$ corresponds to the AA and SP domain wall, respectively. Defining $f = 2 \pi \phi_x / \sqrt{3}$ and $t = y/w$, we obtain
\begin{equation}
    f'' \pm \sin f + \sin 2f = 0,
\end{equation}
with boundary conditions $f(t = \pm \infty) = \pm 2\pi / 3$ and $f(t = \pm \infty) = \mp \pi / 3$, respectively. Here we defined
\begin{equation}
    w = \frac{1}{4\pi} \sqrt{\frac{3}{2c_1}}.
\end{equation}
Multiplying with $2f'$ and integrating gives
\begin{equation} \label{eq:df}
    \left( f' \right)^2 = A \pm 2 \cos f + \cos 2 f,
\end{equation}
where $A = 3/2$ from the boundary conditions. Here we also used that $f'(t = \pm \infty) = 0$. We obtain
\begin{equation}
    f' = \frac{1 \pm 2 \cos f}{\sqrt{2}},
\end{equation}
where the sign is chosen to match the boundary conditions ($f' > 0$ for the AA domain wall and $f' < 0$ for the SP domain wall). The solutions take the form of a Gudermannian function
\begin{align}
    f_\text{AA}(y) = 2 \arctan \left[ \sqrt{3} \tanh \left( \sqrt{\frac{3}{2}} \frac{y}{2w} \right) \right], \\
    f_\text{SP}(y) = -2 \arctan \left[ \frac{1}{\sqrt{3}} \tanh \left( \sqrt{\frac{3}{2}} \frac{y}{2w} \right) \right].
\end{align}
We see that the slope of the AA domain wall is three times larger than that of the SP domain wall. Hence the AA soliton is more narrow than the SP soliton. This is to be expected since the energy barrier is $9$ times larger for AA compared to SP stacking. We further note that our solution for the SP domain wall do not match the solution of Ref.\ \cite{gao_symmetry_2022} and we believe there is a minor arithmetical mistake in that work.  While this does not change their conclusions, we observe that our solution matches the full numerical solution, see Fig.\ \ref{fig:dw}. The full width at half maximum of the soliton is given by (in units $a$)
\begin{align}
    \delta y_\text{AA} & = \frac{\ln(2)}{2\pi\sqrt{c_1}} \approx \frac{0.11}{\sqrt{c_1}}, \\
    \delta y_\text{SP} & = \frac{\ln(1+\sqrt{3})}{2\pi\sqrt{c_1}} \approx \frac{0.16}{\sqrt{c_1}}.
\end{align}
For example, for twisted bilayer graphene using $c_1 \approx 4.5 \times 10^{-5}$ we have $\delta y_\text{SP} \approx 24$ and for twisted WSe$_2$ near $0^\circ$ we have $c_1 \approx 3.9 \times 10^{-4}$ giving $\delta y_\text{SP} \approx 8$  \cite{ezzi_analytical_2024}. However, these values slightly overestimate the domain wall width because of higher-order contributions to the adhesion potential as we discuss below.

We note that Eq.\ \eqref{eq:df} already suffices for our purposes since we are mainly interested in the slope of the domain wall. In this way, we only have to do one integration that remains straightforward when we include higher-order stars in the stacking-fault energy. For the SP domain wall, which is relevant for $D_6$ twist moir\'es, we obtain \cite{ramos-alonso_flat_2025}
\begin{equation} \label{eq:slope}
    \phi_x'(0) = -2 \sqrt{c_1 - 8 c_2 + 9 c_3},
\end{equation}
where $c_{2,3} = V_{2,3}/\mu$ are real coefficients belonging to the second and third star, respectively. For example, we know that for tTMDs, $c_2$ and $c_3$ are generally negative and about ten times smaller in magnitude than $c_1$ \cite{carr_relaxation_2018,bennett_theory_2022}. However, because of the large prefactors in Eq.\ \eqref{eq:slope} these contributions have a significant effect on the soliton size.
Up to the third star, the stacking-fault energy for $C_{6v}$ symmetry is given by
\begin{widetext}
\begin{align}
    V(\bm \phi) & = 2 V_1 \left\{ \cos ( \bm b_1 \cdot \bm \phi ) +  \cos ( \bm b_2 \cdot \bm \phi ) +  \cos [ \left( \bm b_1 + \bm b_2 \right) \cdot \bm \phi ] \right\} \\
    & + 2 V_2 \left\{ \cos [ \left( \bm b_1 + 2 \bm b_2 \right) \cdot \bm \phi ] + \cos [ \left( 2 \bm b_1 + \bm b_2 \right) \cdot \bm \phi ] + \cos [ \left( \bm b_1 - \bm b_2 \right) \cdot \bm \phi ] \right\} \\
    & + 2 V_3 \left\{ \cos ( 2\bm b_1 \cdot \bm \phi ) + \cos ( 2 \bm b_2 \cdot \phi ) + \cos [ 2 \left( \bm b_1 + \bm b_2 \right) \cdot \bm \phi ] \right\}.
\end{align}
\end{widetext}

We show the solution for the isolated single AB/SP/BA domain wall in Fig.\ \ref{fig:dw}. Here we also show the numerical solution of continuum elasticity for the twist moir\'e, as well as the small-angle theory for $V_1 > 0$ that we present in the main text.

\subsection{AA/SP/AA domain wall}

We now consider $V_1 < 0$ and take a domain wall located at $x = 0$ with translational symmetry along $y$ that separates regions with AA stacking (full dislocation). Hence $\bm \phi = \bm \phi(x)$ and the equations of motion become
\begin{equation}
    \begin{bmatrix} \left( \lambda + 2 \mu \right) \phi_x'' \\ \mu \phi_y'' \end{bmatrix} = 2 \frac{\partial V}{\partial \bm \phi},
\end{equation}
where the primes indicate derivatives with respect to the $x$ coordinate. We now set $\phi_x = 0$ and take boundary conditions $f(t = +\infty) = 2\pi$ and $f(t = -\infty) = 0$ with $f = 2\pi \phi_y$. In the first-star approximation, this yields
\begin{equation}
    f' - 2 \sin \tfrac{f}{2} = 0,
\end{equation}
with solution
\begin{equation}
    f(x) = 2 \arccos \left( -\tanh \frac{x}{w'} \right),
\end{equation}
where
\begin{equation}
    w' = \frac{1}{4\pi} \frac{1}{\sqrt{-2c_1}},
\end{equation}
with $c_1 = V_1 / \mu < 0$. The full width at half maximum of the soliton is then given by (in units of $a$)
\begin{equation}
    \delta x = \frac{\cosh^{-1}(3)}{4\pi\sqrt{-2c_1}} \approx \frac{0.1}{\sqrt{-c_1}}.
\end{equation}

Up to the third star, the slope of the AA/SP/AA domain wall is given by
\begin{equation}
    \phi_y'(0) = 4 \sqrt{-2(c_1 + c_2)},
\end{equation}
with $c_2 = V_2/\mu$.

\subsection{Soliton slope of triangular network}

\begin{figure}
    \centering
    \includegraphics[width=\linewidth]{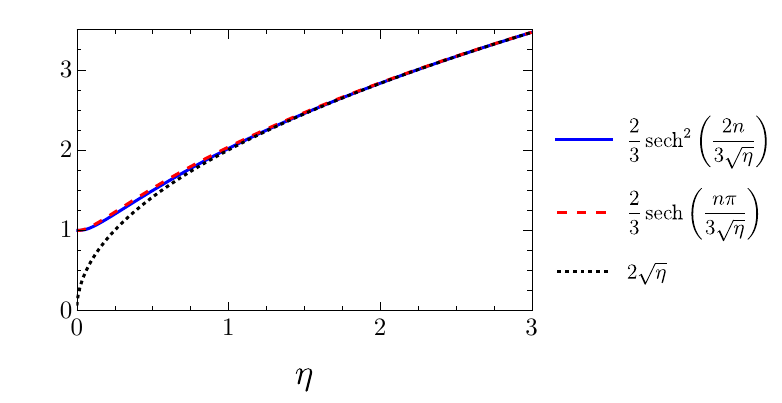}
    \caption{Slope of the triangular soliton network [Eq.\ \eqref{eq:dwseries}] for different parameterizations of $u_{n,0}^\perp$.}
    \label{fig:sm_series}
\end{figure}

In the main text, we constrain the expression for the triangular soliton network by matching the slope of an isolated soliton:
\begin{equation} \label{eq:dwseries}
     2 \sqrt{\eta} = 1 + \sum_{n=1}^\infty \left[ 2 + (-1)^n \right] u^\perp_{n,0}.
\end{equation}
We find this is satisfied for $\eta \gg 1$ by
\begin{equation} \label{eq:sech2}
    u_{n,0}^\perp = \frac{2}{3} \sech^2 \left( \frac{2n}{3\sqrt{\eta}} \right).
\end{equation}
Moreover,
\begin{align}
    u_{n,0}^\perp & = \frac{2}{3} \sech \left( \frac{n\pi}{3\sqrt{\eta}} \right), \\
    u_{n,0}^\perp & = \frac{2}{3} \frac{n\pi^2}{6\sqrt{\eta}} \, \text{csch} \left( \frac{n\pi^2}{6\sqrt{\eta}} \right), \label{eq:csch}
\end{align}
give the same result for the slope, see Fig.\ \ref{fig:sm_series}. Presumably there is a whole family of exponentially decaying functions with similar properties such that they all satisfy Eq.\ \eqref{eq:dwseries} in the limit $\eta \gg 1$. Here Eq.\ \eqref{eq:csch} gives almost the same result as Eq.\ \eqref{eq:sech2} when it come to matching the soliton slope. They agree up to $0.1\%$ so we do not show it in Fig.\ \ref{fig:sm_series}. We opted for Eq.\ \eqref{eq:sech2} because it fitted best to the \textsc{lammps} molcular dynamics simulations for tWSe$_2$.

\section{Numerical solution}

In this section, we derive the self-consistent equations of motion from continuum elasticity, that are solved numerically to obtain the relaxed configuration. We first show that the elastic energy can be written in the form given in Eq.\ (2) of the main text, if we assume that out-of-plane contributions can be neglected. This is justified since the out-of-plane displacements enter at quadratic order in the strain tensor, and are suppressed in encapsulated samples. Hence, in our theory the out-of-plane displacement field is not independent. Instead, it follows the equilibrium value for a given local stacking such that the adhesion potential only depends on the in-plane layer separation.

We denote the in-plane acoustic displacement field of the layers as $\bm u_l$ with $l = 1,2$ the layer index. Assuming an elastically isotropic two-dimensional material, the total elastic energy for a homobilayer can be written as
\begin{equation} \label{eq:elastic}
    \frac{1}{2} \sum_{l=1,2} \int d^2 \bm r \left[ \lambda u_{ii}^{(l)} u_{ii}^{(l)} + 2 \mu u_{ij}^{(l)} u_{ji}^{(l)} \right],
\end{equation}
where $\lambda$ and $\mu$ are in-plane Lam\'e constants, or equivalently, $\lambda + \mu$ is the bulk modulus and $\mu$ is the shear modulus. Here, summation over repeated indices is implied and we introduced
\begin{equation} \label{eq:straintensor}
    u_{ij}^{(l)} = \frac{1}{2} \left( \frac{\partial u_{l,j}}{\partial r_i} + \frac{\partial u_{l,i}}{\partial r_j} \right),
\end{equation}
for $i, j = x, y$. This is the strain tensor relative to the rigid configuration. We now write
\begin{equation}
    \bm u_{1,2}(\bm r) = \overline{\bm u}(\bm r) \pm \bm u(\bm r) / 2,
\end{equation}
with $\overline{\bm u} = ( \bm u_1 + \bm u_2 ) / 2$ and $\bm u = \bm u_1 - \bm u_2$. It follows that the mean and relative displacement fields are decoupled because the cross terms cancel when summing over the layers \cite{ezzi_analytical_2024}.
Hence, Eq.\ \eqref{eq:elastic} can be written as
\begin{equation}
    F_\text{elas}[2\bm u_1] + F_\text{elas}[2\bm u_2] = F_\text{elas}[2\overline{\bm u}] + F_\text{elas}[\bm u],
\end{equation}
with
\begin{equation}
    F_\text{elas}[\bm u] = \frac{1}{4} \int d^2 \bm r \left[ \lambda u_{ii} u_{ii} + 2 \mu u_{ij} u_{ji} \right].
\end{equation}
Furthermore, since the interlayer disregistry
\begin{equation}
    \phi_i = M_{ij} r_j + u_i,
\end{equation}
we have
\begin{equation}
    \phi_{ij} = \frac{1}{2} \left( \partial_i \phi_j + \partial_j \phi_i \right) = \frac{M_{ji} + M_{ij}}{2} + u_{ij},
\end{equation}
and therefore the elastic energy can be written in terms of $\bm \phi(\bm r)$ up to constants and boundary terms. The latter vanish for a periodic solution. For example, terms of the form $M_{ii} M_{ii}$ and $M_{ij} u_{ji}$ in the integrand. For a twist moir\'e, $M_{ij} = -M_{ji}$ and these terms vanish regardless of boundary conditions. This makes sense, as a rigid twist does not contribute to the elastic energy. Hence, in general we find that
\begin{equation}
    F_\text{elas}[\bm u] = F_\text{elas}[\bm \phi] + \text{constant}.
\end{equation}

Under the assumptions outlined above, we find that the elastic and adhesion energy of a homobilayer moir\'e material can be written as
\begin{align}
    F_\text{elas} & = \int d^2 \bm r \left[ \frac{\lambda}{4} \left( \partial_i \phi_i \right)^2 + \frac{\mu}{8} \left( \partial_i \phi_j + \partial_j \phi_i \right)^2 \right], \\
    F_\text{adh} & = \int d^2 \bm r \, V[\bm \phi(\bm r)],
\end{align}
where $\bm \phi = M \bm r + \bm u$ is the interlayer disregistry. Note that $\overline{\bm u}$ is constant (we set it to zero) because the adhesion energy only depends on $\bm u = \bm u_1 - \bm u_2$. 
Specifically, for a twist homobilayer,
\begin{equation}
    \bm \phi(\bm r) = \frac{a}{L} \, \hat z \times \bm r + \bm u(\bm r),
\end{equation}
where
\begin{equation}
    \bm u(\bm r) = \sum_{\bm g} \bm u_{\bm g} e^{i\bm g \cdot \bm r},
\end{equation}
gives the acoustic displacement field with $\bm g$ moir\'e reciprocal lattice vectors. We imposed moir\'e periodic boundary conditions and we set $\bm u_{\bm g = \bm 0} = \bm 0$ since it corresponds to an overall translation of the moir\'e lattice in the long-wavelength limit. We further have
\begin{equation}
    \bm b \cdot \bm \phi(\bm r) = \bm g \cdot \bm r + \bm b \cdot \bm u(\bm r),
\end{equation}
where 
$\bm b$ is a reciprocal lattice vector of the monolayer with $\bm g = M^\top \bm b = (a / L) \bm b \times \hat z$. The total energy density $f = f_\text{elas} + f_\text{adh}$ then becomes
\begin{align}
    f_\text{elas} & = \frac{a^2}{4L^2} \sum_{\bm g} \left[ \left( \lambda + 2 \mu \right) | u_{\bm g}^\parallel |^2 + \mu | u_{\bm g}^\perp |^2 \right], \\
    f_\text{adh} & = \frac{1}{A} \int_\text{moir\'e cell} d^2 \bm r \, V[\bm \phi(\bm r)], \label{eq:fadh}
\end{align}
where $A = | \bm L_1 \times \bm L_2 |$ is the moir\'e cell area and
\begin{equation}
    u_{\bm g}^\parallel = \frac{iL}{a} \bm g \cdot \bm u_{\bm g}, \qquad u_{\bm g}^\perp = \frac{iL}{a} \left( \hat z \times \bm g \right) \cdot \bm u_{\bm g}, 
\end{equation}
are (scaled) longitudinal and transverse components of the displacement field. Here we anticipate writing $\bm g$ and $\bm u$ in units of $1/L$ and $a$, respectively.
\begin{figure}
    \centering
    \includegraphics[width=.6\linewidth]{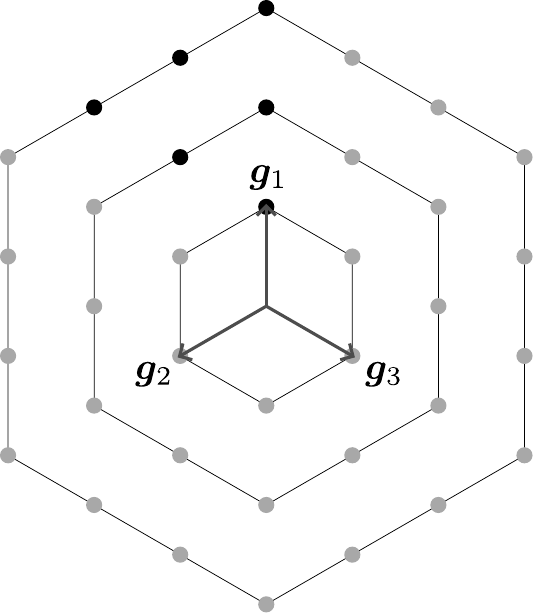}
    \caption{The first three reciprocal shells where we omit the zeroth shell, i.e., the origin. Each shell corresponds to the reciprocal vectors (points) that lie on a hexagon. These are further divided into reciprocal stars that are closed under $60^\circ$ rotations: the $n$th shell contains $n$ stars. Here the black dots give one choice of representatives, while the gray dots give reciprocal vectors related by symmetry.}
    \label{fig:shells}
\end{figure}

\subsection{Self-consistent solution}

The gradient of the energy with respect to the components is given by
\begin{align}
    \frac{\partial f}{\partial u_{-\bm g}^\parallel} & = \frac{a^2}{2L^2} \left( \lambda + 2 \mu \right) u_{\bm g}^\parallel + \frac{\partial f_\text{adh}}{\partial u_{-\bm g}^\parallel}, \\
    \frac{\partial f}{\partial u_{-\bm g}^\perp} & = \frac{a^2}{2L^2} \mu u_{\bm g}^\perp + \frac{\partial f_\text{adh}}{\partial u_{-\bm g}^\perp},
\end{align}
with
\begin{align}
    \frac{\partial f_\text{adh}}{\partial u_{-\bm g}^\parallel} & = \frac{1}{A} \int_\text{moir\'e cell} d^2 \bm r \, \frac{\partial \bm u}{\partial u_{-\bm g}^\parallel} \cdot \frac{\partial V}{\partial \bm \phi} \\
    & = -\frac{a}{L} \frac{\bm g}{ig^2} \cdot 
    \left( \frac{\partial V}{\partial \bm \phi} \right)_{\bm g}, \\
    \frac{\partial f_\text{adh}}{\partial u_{-\bm g}^\perp}  & = - \frac{a}{L} \frac{\hat z \times \bm g}{ig^2} \cdot \left( \frac{\partial V}{\partial \bm \phi} \right)_{\bm g},
\end{align}
with
\begin{equation}
    \frac{\partial V}{\partial \bm \phi} = \sum_{\bm b} i \bm b V_{\bm b} e^{i \bm b \cdot \bm \phi}.
\end{equation}

Hence, the self-consistency equations are given by
\begin{align}
    u_{\bm g}^\parallel & = \frac{2L^2}{a^2} \frac{\bm g}{Lg^2} \cdot \left( \frac{-ia}{\lambda + 2 \mu} \frac{\partial V}{\partial \bm \phi} \right)_{\bm g}, \label{eq:scpara} \\
    u_{\bm g}^\perp & = \frac{2L^2}{a^2} \frac{\hat z \times \bm g}{Lg^2} \cdot \left( \frac{-ia}{\mu} \frac{\partial V}{\partial \bm \phi} \right)_{\bm g}. \label{eq:scperp}
\end{align}
In this work, we only consider moir\'es with $\mathcal C_{3z}$ symmetry such that we only need to solve for two complex coefficients $u_{n,m}^\parallel$ and $u_{n,m}^\perp$ for each reciprocal star. We define a star as a collection of six reciprocal vectors of equal length that are closed under $\mathcal C_{6z}$ rotations. The stars are grouped into shells where the $n$th reciprocal shell contains a total of $n$ stars with $m=0,\ldots,n-1$. These correspond to the collection of reciprocal vectors that lie on the edges of a hexagon of radius $4\pi n/\sqrt{3}L$. This is illustrated in Fig.\ \ref{fig:shells}. Hence, we only need to choose one representative
$\bm g_{n,m} = n\bm g_1 + m \bm g_2$ for each star. The self-consistency equations are independent of the choice of representative because 
\begin{equation}
    \left( \frac{\partial V}{\partial \bm \phi} \right)_{\mathcal S \bm g} = \mathcal S \left( \frac{\partial V}{\partial \bm \phi} \right)_{\bm g},
\end{equation}
for a rotation symmetry $\mathcal S$. The displacement field is then given by
\begin{equation}
    \begin{aligned}
        \frac{\bm u(\bm r)}{a} & = 2 \text{Re} \sum_{n=1}^\infty \sum_{m=0}^{n-1} \sum_{j=1}^3 \\
        & \times \mathcal C_{3z}^j \frac{u_{n,m}^\parallel \bm g_{n,m} + u_{n,m}^\perp \hat z \times \bm g_{n,m}}{i L g_{n,m}^2} \, e^{i \mathcal C_{3z}^j \bm g_{n,m} \cdot \bm r}.
    \end{aligned}
\end{equation}
The parameters of this theory are given by $\lambda/\mu$ and $\eta_i = (L^2/a^2) V_i/\mu$.

\subsection{Perturbation theory}

In lowest order of $|\bm u(\bm r)/a|$ we can set the displacement field to zero on the right-hand side of Eqs.\ \eqref{eq:scpara} and \eqref{eq:scperp}. This yields the one-shot result:
\begin{align}
    u_{\bm g}^\parallel & \simeq \frac{2L^2}{a^2} \frac{\bm g \cdot \bm b}{Lg^2} \frac{a V_{\bm g}}{\lambda + 2 \mu}, \\
    u_{\bm g}^\perp & \simeq \frac{2L^2}{a^2} \frac{( \hat z \times \bm g ) \cdot \bm b}{Lg^2} \frac{aV_{\bm g}}{\mu},
\end{align}
with $\bm b = \tfrac{L}{a} \hat z \times \bm g$ such that in lowest order
\begin{equation}
    u_{\bm g}^\parallel = 0, \qquad
    u_{\bm g}^\perp = \frac{2L^2}{a^2} \frac{V_{\bm g}}{\mu} = 2 \eta_{\bm g}.
\end{equation}

Alternatively, we can expand the adhesion energy \ref{eq:fadh} up to first-order in the displacement field \cite{ezzi_analytical_2024}. This is allows for a more systematic perturbative expansion. Noting that
\begin{equation}
    V[\bm\phi(\bm r)] = \sum_{\bm g} V_{\bm g} e^{i\bm g \cdot \bm r} e^{i \bm g \cdot M^{-1}\bm u(\bm r)},
\end{equation}
we have
\begin{widetext}
\begin{align}
    f_\text{adh} & = \frac{1}{A} \sum_{\bm g} V_{\bm g} \int d^2\bm r \, e^{i\bm g \cdot \bm r} \left\{ 1 +  i \bm g \cdot M^{-1} \bm u(\bm r) - \frac{1}{2} \left[ \bm g \cdot M^{-1} \bm u(\bm r) \right]^2 + \cdots \right\} \\
    & = V_0 + \sum_{\bm g} V_{\bm g} \left( i \bm g \cdot M^{-1} \bm u_{-\bm g} \right) - \frac{1}{2} \sum_{\bm g} V_{\bm g} \sum_{\bm g' \neq \bm g} \left( \bm g \cdot M^{-1} \bm u_{-\bm g'} \right) \left( \bm g \cdot M^{-1} \bm u_{\bm g'-\bm g} \right) + \cdots.
\end{align}
We can systematically perform this expansion given an \emph{ansatz} for $\bm u(\bm r)$ by simply expanding the integrand in Mathematica and selecting terms that do not depend on the coordinates. For twist moir\'es, we have $M^{-1} \bm u_{\bm g} = (L/a) \bm u_{\bm g} \times \hat z$ and we obtain
\begin{equation}
    f_\text{adh} = V_0 - \sum_{\bm g} V_{\bm g} u_{-\bm g}^\perp - \frac{L^2}{2a^2} \sum_{\bm g} V_{\bm g} \sum_{\bm g' \neq \bm g} \left( \hat z \times \bm g \cdot \bm u_{-\bm g'} \right) \left( \hat z \times \bm g \cdot \bm u_{\bm g'-\bm g} \right) + \cdots.
\end{equation}
\end{widetext}
One could also have used the Jacobi-Anger identity,
\begin{align}
    V[\bm\phi(\bm r)] & = \sum_{\bm g} V_{\bm g} e^{i\bm g \cdot \bm r} e^{i \bm g \cdot M^{-1} \sum_{\bm g'} i\bm u_{\bm g'} \sin(\bm g'\cdot \bm r)} \\
    & = \sum_{\bm g} V_{\bm g} e^{i\bm g\cdot \bm r}\nonumber \\
    & \times \prod_{\bm g'} \sum_{m=-\infty}^\infty J_m(\bm g \cdot M^{-1} i\bm u_{\bm g'}) e^{im\bm g'\cdot \bm r},
\end{align}
to expand the adhesion energy in an infinite series of Bessel functions \cite{ceferino_pseudomagnetic_2024,kang_analytical_2025}. However, since in the end analytical progress can only be made by expanding the adhesion energy, we prefer to directly expand it in powers of $|\bm u/a|$.
\begin{figure}
    \centering
    \includegraphics[width=\linewidth]{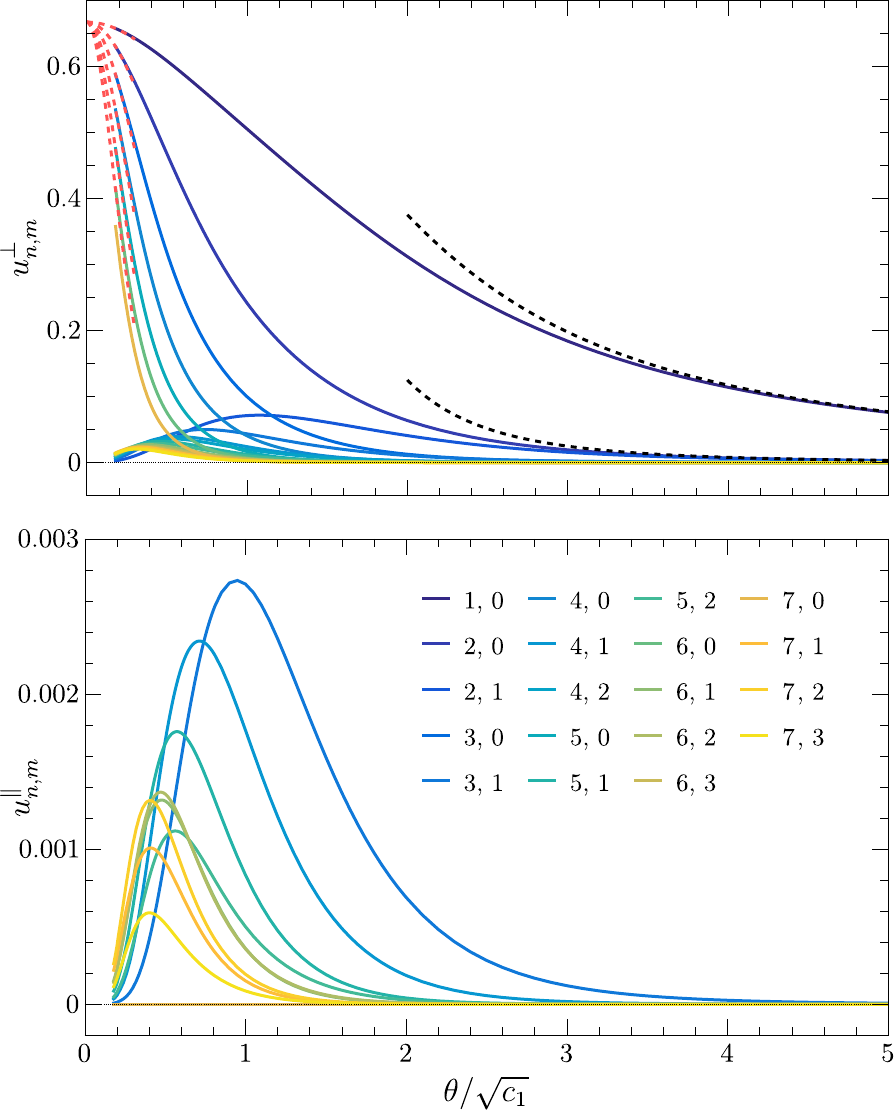}
    \caption{Numerical solution for $V_1>0$ as a function of $1/\sqrt{\eta}$. We show distinct values not related by symmetry for the first $7$ shells ($28$ stars). Note that the longitudinal components are more than two orders of magnitude smaller and are maximal in the intermediate regime $\eta \approx 1$. For the transverse components, the dashed lines give the perturbative result for ``large" angles (black) and the \emph{ansatz} for ``small" angles.}
    \label{fig:sol0}
\end{figure}

For example, if we only include the first three stars, we only have to consider three transverse components $u_1 = u^\perp_{1,0}$, $u_2 = u^\perp_{2,1}$, and $u_3 = u^\perp_{2,0}$ because the longitudinal components are forbidden by $D_6$ symmetry. Up to third order, including also three stars for the adhesion potential, we obtain
\begin{equation}
    f_\text{elas} = \frac{3\mu}{2} \frac{a^2}{L^2} \left( u_1^2 + u_2^2 + u_3^2 \right), 
\end{equation}
and (ignoring the constant term $V_0$) up to third order in $|\bm u / a|$ we find
\begin{widetext}
\begin{align}
    f_\text{adh} & = V_1 \left\{ \frac{3}{8} \left[ 8 u_1 \left( 2u_1^2 + u_1 u_2 + u_2^2 \right) + 2u_2u_3 \left( 3u_1 + 2u_2 \right) + 7u_1 u_3^2 \right] + \frac{3}{2} u_1 \left[ u_1 - 2 \left( u_2 + u_3 \right) \right] - 6 u_1 \right\} \\
    & + V_2 \left\{ \frac{3}{8} \left[ 16 u_2^3 + 36 u_1 u_2 u_3 + 27 u_2 u_3^2 + 18 u_1^2 \left( 4u_2 + 3 u_3 \right) \right] + \frac{3}{2} \left( 9u_1^2 + u_2^2 \right) - 6 u_2 \right\} \\
    & + V_3 \left\{ 3 \left[ -4u_1^3 + 5 u_2^2 u_3 + 2 u_3^3 + 2u_1 u_2 \left( 2u_2 + u_3 \right) + u_1^2 \left( 8u_2 + 11 u_3 \right) \right] + \frac{3}{2} \left[ 8u_1 \left( u_1 + u_2 \right) + u_3^2 \right] - 6 u_3 \right\}.
\end{align}
\end{widetext}
This then yields three equations ($i=1,2,3$)
\begin{equation}
    \frac{\partial}{\partial u_i} \left( f_\text{elas} + f_\text{adh} \right) = 0,
\end{equation}
which are solved perturbatively using the Frobenius method by expanding each $u_i$ in powers of $\eta_j = (L^2/a^2) V_j/\mu \approx V_j / ( \theta^2 \mu)$ ($j=1,2,3$). We find
\begin{align}
    u_1 & \approx 2 \eta_1 - 2 \left[ \eta_1 \left( \eta_1 + 8 \eta_2 + 7 \eta_3 \right) + 4\eta_2 \eta_3 \right], \\
    u_2 & \approx 2 \eta_2 + 2 \left( \eta_1^2 - \eta_2^2 - 4\eta_1 \eta_3 \right),  \\
    u_3 & \approx 2 \eta_3 + 2 \left( \eta_1^2 - \eta_3^2 \right),
\end{align}
or
\begin{align}
    u_{1,0}^\perp & \approx \frac{2c_1}{\theta^2} - \frac{2\left[ c_1 \left( c_1 + 8 c_2 + 7 c_3 \right) + 4c_2 c_3 \right]}{\theta^4}, \\
    u_{2,1}^\perp & \approx \frac{2c_2}{\theta^2} + \frac{2 \left( c_1^2 - c_2^2 - 4c_1c_3 \right)}{\theta^4},  \\
    u_{2,0}^\perp & \approx \frac{2c_3}{\theta^2} + \frac{2 \left( c_1^2 - c_3^2 \right)}{\theta^4}.
\end{align}
Here the dimensionless constants $c_i = V_i / \mu$ can be obtained by fitting the large twist angle scaling behavior to \emph{ab initio} methods \cite{ezzi_analytical_2024}. We compare our perturbative result to a recent work \cite{kang_analytical_2025}. We find that by going to third order for all three stars, we obtain $2\alpha = 1 + 8c_2/c_1 + 7c_3/c_1 \approx 0.05$ using the parameters for twisted bilayer graphene of Ref.\ \cite{carr_relaxation_2018}.
\begin{figure}
    \centering
    \includegraphics[width=\linewidth]{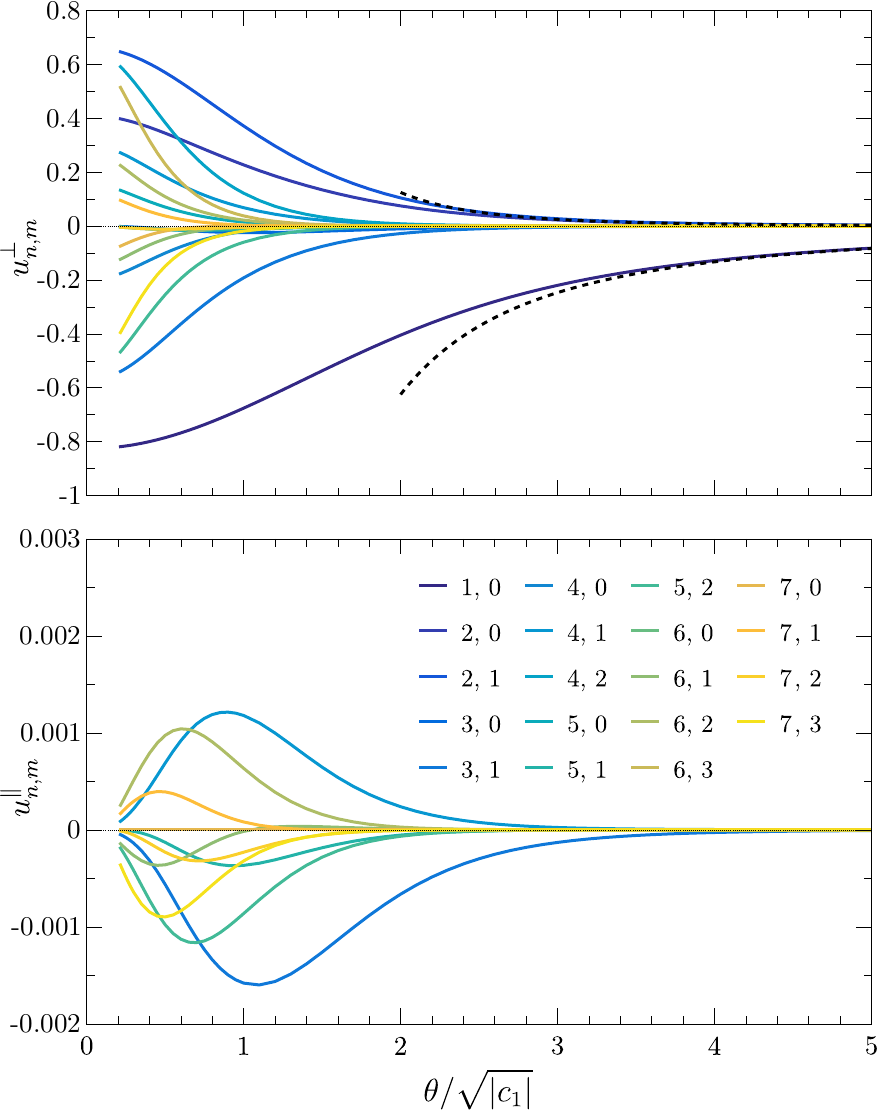}
    \caption{Numerical solution for $V_1<0$ as a function of $1/\sqrt{|\eta|}$. We show distinct values not related by symmetry for the first $7$ shells ($28$ stars). Note that the longitudinal components are more than two orders of magnitude smaller and are maximal in the intermediate regime $\eta \approx -1$.}
    \label{fig:sol60}
\end{figure}
\begin{figure*}
    \centering
    \includegraphics[width=.95\linewidth]{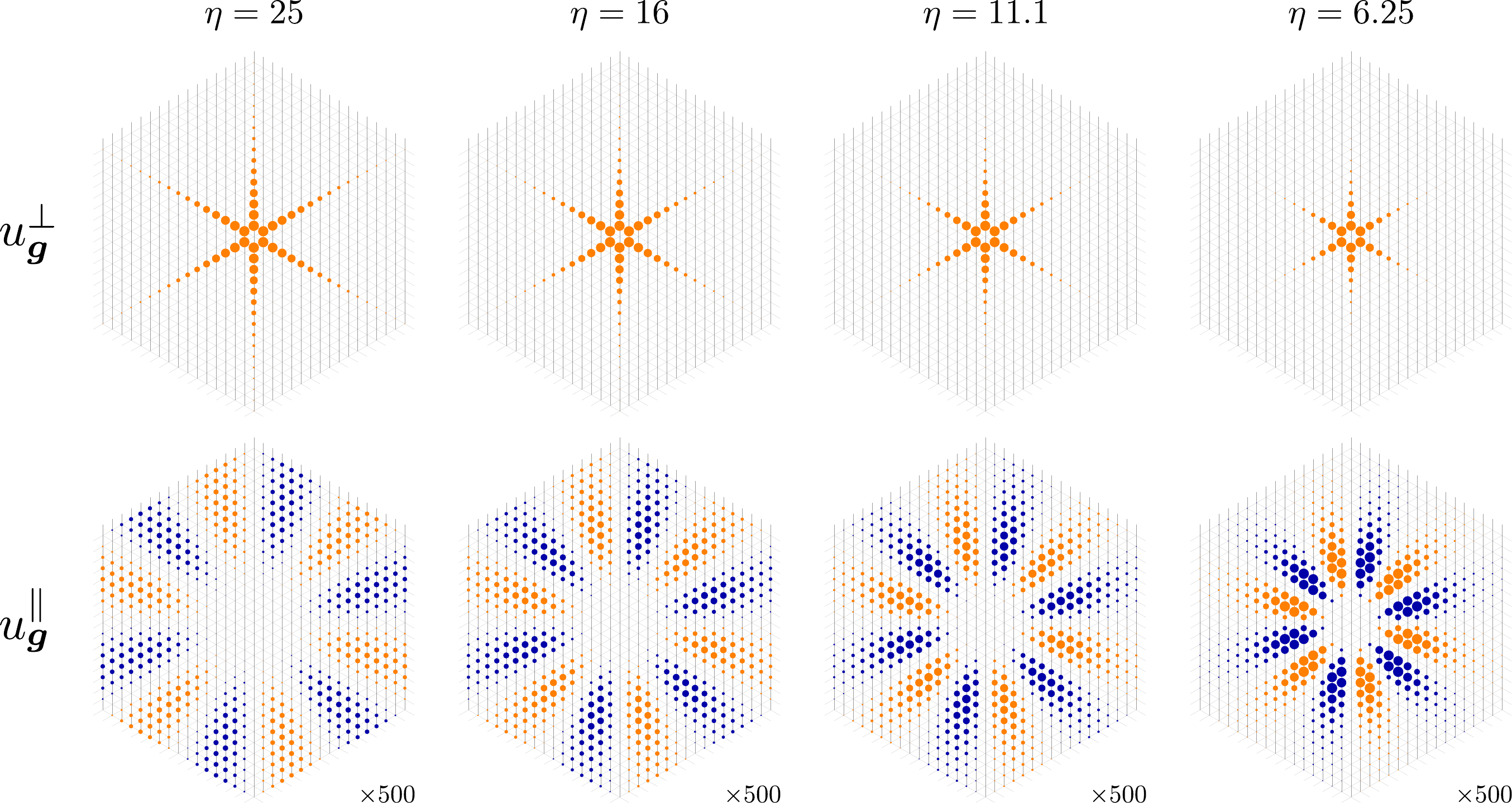}
    \caption{Solution of continuum elasticity for a $D_6$ twist moir\'e in the first-star approximation with $V_1 > 0$ and $\lambda = 0$, showing $16$ shells ($136$ stars). Here the dot size gives the magnitude of the Fourier components and the color indicates the sign, where orange (blue) is positive (negative). The longitudinal components $u_{\bm g}^\parallel$ are scaled by a factor $500$.}
    \label{fig:sm_blob_0}
\end{figure*}
\begin{figure*}
    \centering
    \includegraphics[width=.95\linewidth]{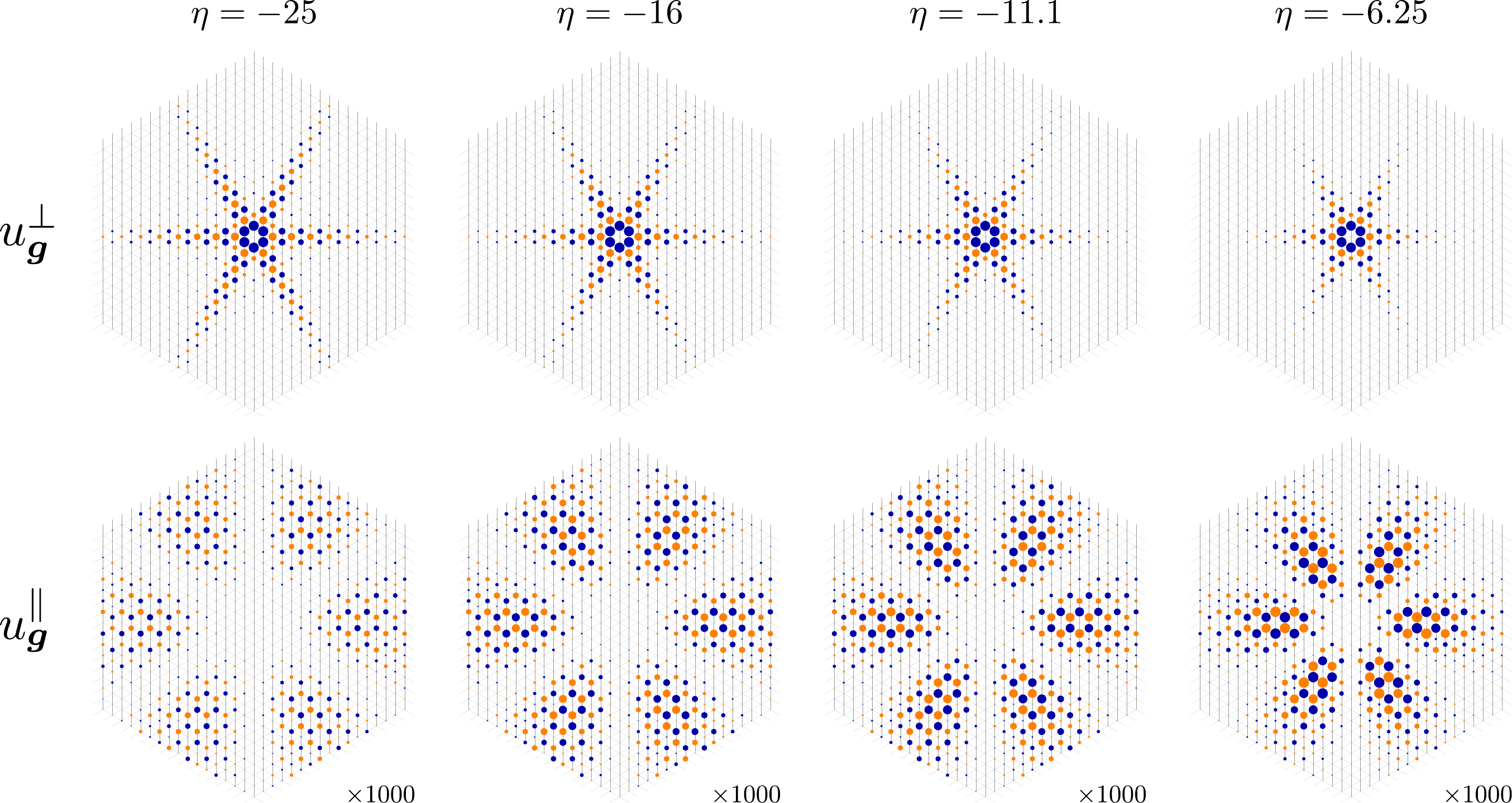}
    \caption{Solution of continuum elasticity for a $D_6$ twist moir\'e in the first-star approximation with $V_1 < 0$ and $\lambda = 0$, showing $16$ shells ($136$ stars). Here the dot size gives the magnitude of the Fourier components and the color indicates the sign, where orange (blue) is positive (negative). The longitudinal components $u_{\bm g}^\parallel$ are scaled by a factor $1000$.}
    \label{fig:sm_blob_60}
\end{figure*}
\begin{table}
    \label{tab:D6}
    \centering
    \begin{tabular}{c | c | c }
        \Xhline{1pt}
        $n,m$ & $u_{n,m}^\parallel$ & $u_{n,m}^\perp$ \\
        \hline
        $1,0$ & $0$ & $\mathds R$ \\ 
        $2,1$ & $0$ & $\mathds R$ \\
        $2,0$ & $0$ & $\mathds R$ \\
        $3,1$ & $\mathds R$ & $\mathds R$ \\
        $3,2$ & $-u_{3,1}^\parallel$ & $u_{3,1}^\perp$ \\
        \Xhline{1pt}
    \end{tabular} \quad
    \begin{tabular}{c | c | c }
        \Xhline{1pt}
        $n,m$ & $u_{n,m}^\parallel$ & $u_{n,m}^\perp$ \\
        \hline
        $1,0$ & $0$ & $\mathds C$ \\ 
        $2,1$ & $i\mathds R$ & $\mathds R$ \\
        $2,0$ & $0$ & $\mathds C$ \\
        $3,1$ & $\mathds C$ & $\mathds C$ \\
        $3,2$ & $-(u_{3,1}^\parallel)^*$ & $(u_{3,1}^\perp)^*$ \\
        \Xhline{1pt}
    \end{tabular}
    \caption{Symmetry-allowed values for the in-plane and out-of-plane Fourier coefficients of the displacement fields in the presence of $D_6$ (left) and $D_3 = \left< \mathcal C_{3z}, \mathcal C_{2y} \right>$ (right) symmetry for the first five reciprocal stars.}
\end{table}

\subsection{First-star theory}

If we restrict the stacking-fault energy to the first star, then it only depends on two real parameters $c=c_1$ and $\psi=\psi_1$. In this case, we define $\eta = \sqrt{c} L / a$. We see that for $\eta \ll 1$ we obtain Eq.\ (3) of the main text with $u_m^\perp = 2\eta^2 e^{i\psi} \delta_{m1}$ and $u_m^\parallel = 0$. For $\eta \gtrsim 1$, we solve the self-consistency equations numerically. Writing $\bm g$ in units of $1/L$, $\partial V/\partial \bm \phi$ in units of $V_1/a$, and $\lambda$ in units of $\mu$, we have
\begin{align}
    u_{\bm g}^\parallel & = 2 \eta^2 \frac{\bm g}{g^2} \cdot \left( \frac{-i}{\lambda + 2} \frac{\partial V}{\partial \bm \phi} \right)_{\bm g}, \\
    u_{\bm g}^\perp & = 2 \eta^2 \frac{\hat z \times \bm g}{g^2} \cdot \left( -i \frac{\partial V}{\partial \bm \phi} \right)_{\bm g}.
\end{align}
Some results for large $\eta$ are shown in Fig.\ \ref{fig:sm_blob_0}. These results are all consistent with a symmetry analysis for $D_6$ twist moir\'es \cite{ezzi_analytical_2024} shown in Table \ref{tab:D6}. Longitudinal components $u_{\bm g}^\parallel$ are only allowed in pairs of opposites corresponding to two stars that map into each other under $\mathcal C_{2x}$. Hence they vanish for any star that is closed under this symmetry. We also see that the magnitude of $u_{\bm g}^\parallel$ decays with increasing $\eta$. Moreover, the largest $|u_{\bm g}^\parallel|$ is almost three orders of magnitude smaller than the largest $|u_{\bm g}^\perp|$.

If we consider two different materials with different $c$ then we expect the same relaxation physics for
\begin{equation}
    \frac{\theta'}{\theta} = \sqrt{\frac{c'}{c}}.
\end{equation}
For example, for group VI twisted bilayer 2H TMDs such as tWSe$_2$ and tMoTe$_2$: $c \sim 10 \times c_\text{tBG}$ such that we expect similar moir\'e reconstruction as compared to tBG for twist angles that are about a factor $3$ larger \cite{carr_relaxation_2018,bennett_theory_2022,ezzi_analytical_2024}.

\subsubsection{Numerical implementation}

We solve the self-consistency equations numerically in \textsc{julia}. In the numerical implementation we take a finite the number of shells $N$ giving a total number of stars $M = N(N+1)/2$ with representatives $\bm g_m$ ($m = 1,\ldots, M$). A configuration can then be defined by the longitudinal $U^\parallel$ and transverse $U^\perp$ components with
\begin{equation}
    U^\parallel = \begin{pmatrix} u_1^\parallel \\ u_2^\parallel \\ \vdots \\ u_M^\parallel \end{pmatrix}, \qquad U^\perp = \begin{pmatrix} u_1^\perp \\ u_2^\perp \\ \vdots \\ u_M^\perp \end{pmatrix},
\end{equation}
from which one can construct $\bm u(\bm r)$. In fact, one can roughly half the number of stars further by taking into account $\mathcal C_{2x}$ symmetry, see for example Table \ref{tab:D6} for the fourth and fifth star. However, in practice we only take into account $\mathcal C_{6z}$ symmetry which makes all coefficients real. We proceed by calculating the right-hand side of Eq.\ \eqref{eq:scpara} and Eq.\ \eqref{eq:scperp} numerically by computing $M$ integrals in parallel. To avoid convergence issues, instead of directly updating the $u_m^\parallel$ and $u_m^\perp$, we use the \textsc{diis} method (direct inversion of the iterative subspace) which was originally developed for the self-consistent field method \cite{noauthor_diis_nodate,pulay_improved_1982}. To further stabilize the algorithm, we always start in the perturbative regime $\eta \ll 1$ (large twist angles) where the exact solution is known and the solution converges quickly. Since the solution should be continuous as a function of $\eta$ (or the twist angle), we gradually increase $\eta$ (lower the twist angle), using the converged result of the previous larger angle as the starting point for the new smaller angle. Similarly, one can increase the number of shells $N$ between consecutive runs. For our purposes we use $N = 18$. To make sure the results are converged properly, we always check whether the constraints from $\mathcal C_{2x}$ symmetry are satisfied.

\section{Heterostrain}

In this section, we briefly discuss how our numerical approach can be extended to include heterostrain. Following Ref.\ \cite{kang_analytical_2025} we use the original reciprocal vectors $\bm b$ of the monolayers to label the Fourier components. This is more convenient as these are always defined on the same hexagonal lattice in reciprocal space. The adhesion potential is
\begin{equation}
    V(\bm \phi) = \sum_{\bm b} e^{i \bm b \cdot \bm \phi} V_{\bm b},
\end{equation}
In the bilayer moir\'e cell, the local stacking is
\begin{equation}
    \bm \phi(\bm r) = M \bm r + \bm u(\bm r),
\end{equation}
with $\bm r \in s_1 \bm L_1 + s_2 \bm L_2$ where $s_{1,2} \in [0,1)$, and $\bm u = \bm u_1 - \bm u_2$ is the acoustic displacement field with
\begin{equation}
    \bm u(\bm r) = \sum_{\bm g} \bm u_{\bm g} e^{i \bm g \cdot \bm r} = \sum_{\bm b} \bm u_{\bm b} e^{i \bm b \cdot M \bm r} \equiv \bm U(M \bm r).
\end{equation}
We can further define
\begin{equation}
    \bm \Phi(\bm r) \equiv \bm \phi(M^{-1} \bm r) = \bm r + \bm U(\bm r),
\end{equation}
with $\bm r \in s_1 \bm a_1 + s_2 \bm a_2$. Here, we can either label the Fourier components with the monolayer or moir\'e reciprocal vectors. As mentioned above, it will be more convenient to use the monolayer vectors because they are always defined on an undistorted hexagonal grid. The moir\'e lattice is then defined by $\bm \phi(\bm r + \bm L) = \bm \phi(\bm r) + \bm a \equiv \bm \phi(\bm r)$ which yields
\begin{equation}
    \bm g = M^\top \bm b, \qquad \bm L = M^{-1} \bm a.
\end{equation}
In general, the rigid moir\'e is defined by the constant displacement gradient
\begin{equation}
    M = \left( 1 + \frac{\mathcal E}{2} \right) R(\theta/2) - \left( 1 - \frac{\mathcal E}{2} \right) R(-\theta/2) \approx \mathcal E - i \theta \sigma_y,
\end{equation}
where $\mathcal E$ describes both lattice mismatch and external heterostrain.

Similar as before, the elastic and adhesion energies are given by
\begin{align}
    F_\text{elas}[\bm \phi] & = \int d^2 \bm r \left[ \frac{\lambda}{4} \left( \partial_i \phi_i \right)^2 + \frac{\mu}{8} \left( \partial_i \phi_j + \partial_j \phi_i \right)^2 \right], \label{eq:Felas} \\
    F_\text{adh}[\bm \phi] & = \int d^2 \bm r \, V[\bm \phi(\bm r)],
\end{align}
where
\begin{equation}
    \partial_j \phi_i = M_{ij} + \partial_j u_i.
\end{equation}
Therefore, we can write
\begin{widetext}
\begin{align}
    F_\text{elas} & = \int d^2 \bm r \left[ \frac{\lambda}{4} \left( M_{ii} + \partial_i u_i \right)^2 + \frac{\mu}{8} \left( M_{ij} + M_{ji} + \partial_i u_j + \partial_j u_i \right)^2 \right] \\
    & = \int d^2 \bm r \left[ \frac{\lambda}{4} \left( \partial_i u_i \right)^2 + \frac{\mu}{8} \left( \partial_i u_j + \partial_j u_i \right)^2 \right],
\end{align}
where we discarded constants and boundary terms. After performing a Fourier transform on the displacement fields, the elastic energy density becomes
\begin{align}
    f_\text{elas} & = -\sum_{\bm g, \bm g'} \frac{1}{A_\mathrm{m}} \int_\text{moir\'e cell} d^2 \bm r \left[ \frac{\lambda}{4} ( \bm g \cdot \bm u_{\bm g} ) ( \bm g' \cdot \bm u_{\bm g'} ) + \frac{\mu}{8} \left( g_i u_{\bm g,j} + g_j u_{\bm g,i} \right) \left( g_i' u_{\bm g',j} + g_j' u_{\bm g',i} \right) \right] e^{i(\bm g+\bm g')\cdot \bm r} \\
    & = \frac{1}{4} \sum_{\bm g} \left[ \lambda | \bm g \cdot \bm u_{\bm g} |^2 + \frac{\mu}{2} \left| g_i u_{\bm g,j} + g_j u_{\bm g,i} \right|^2 \right] \\
    & = \frac{1}{4} \sum_{\bm g} \left[ \left( \lambda + 2 \mu \right) | \bm g \cdot \bm u_{\bm g} |^2 + \mu \left| (\hat z \times \bm g) \cdot \bm u_{\bm g} \right|^2 \right].
\end{align}
\end{widetext}
In turn, this can be written as
\begin{equation}
    f_\mathrm{elas} = \frac{a^2}{4L^2} \sum_{\bm g} \left[ \left( \lambda + 2 \mu \right) | u_{\bm g}^\parallel |^2 + \mu | u_{\bm g}^\perp |^2 \right],
\end{equation}
where we defined $L = (L_1+L_2)/2$ for the general case with $L_1 = |\bm L_1| \neq |\bm L_2|=L_2$. Here we used a Helmholtz decomposition, which is always possible for a vector field on a torus,
\begin{equation}
    \bm u_{\bm g} = \frac{a}{L} \frac{u_{\bm g}^\parallel \bm g + u_{\bm g}^\perp \hat z \times \bm g}{ig^2}.
\end{equation}
We also have
\begin{align}
    f_\mathrm{adh} & = \frac{1}{A_\mathrm{m}} \int_\text{moir\'e cell} d^2\bm r \, V[\bm \phi(\bm r)] \\
    & = \sum_{\bm b} V_{\bm b} \int_0^1 ds_1 \int_0^1 ds_2 \\
    & \qquad \times e^{i \bm b \cdot \left[ s_1 \bm a_1 + s_2 \bm a_2 + \bm U(s_1 \bm a_1 + s_2 \bm a_2) \right]}, \nonumber
\end{align}
where we used
\begin{align}
    \bm u(s_1 \bm L_1 + s_2 \bm L_2) & = \sum_{\bm b} \bm u_{\bm b} e^{i\bm b \cdot (s_1 \bm a_1 + s_2 \bm a_2)} \\
    & = \bm U(s_1 \bm a_1 + s_2 \bm a_2).
\end{align}
For a general rigid moir\'e and an even adhesion potential $V(\bm \phi) = V(-\bm \phi)$, the displacement field only has the following symmetry:
\begin{equation}
    \bm u(-\bm r) = -\bm u(\bm r) = 2 \sideset{}{'}\sum_{\bm b} u_{\bm b} \sin(\bm b \cdot M \bm r),
\end{equation}
where the primed sum only runs over half of the reciprocal vectors, excluding zero. 

To minimize the energy, we require
\begin{widetext}
\begin{align}
    \frac{\partial f}{\partial u_{-\bm g}^\parallel} & = \frac{a^2}{2L^2} (\lambda + 2 \mu) u_{\bm g}^\parallel + \frac{1}{A_\mathrm{m}} \int_\text{moir\'e cell} d^2\bm r \, \frac{\partial \bm u}{\partial u_{-\bm g}^\parallel} \cdot \frac{\partial V}{\partial \bm \phi}, \\
    \frac{\partial f}{\partial u_{-\bm g}^\perp} & = \frac{a^2}{2L^2} \mu u_{\bm g}^\perp + \frac{1}{A_\mathrm{m}} \int_\text{moir\'e cell} d^2\bm r \, \frac{\partial \bm u}{\partial u_{-\bm g}^\perp} \cdot \frac{\partial V}{\partial \bm \phi},
\end{align}
with
\begin{align}
    \frac{\partial V}{\partial \bm \phi} & = i \sum_{\bm b} \bm b e^{i\bm b \cdot \bm \phi} V_{\bm b}, \\
    \frac{\partial \bm u}{\partial u_{-\bm g}^\parallel} & = -\frac{a}{L} \frac{\bm g}{ig^2} e^{-i \bm g \cdot \bm r}, \\
    \frac{\partial \bm u}{\partial u_{-\bm g}^\perp} & = -\frac{a}{L} \frac{\hat z \times \bm g}{ig^2} e^{-i \bm g \cdot \bm r}.
\end{align}
We thus obtain
\begin{align}
    \frac{\partial f}{\partial u_{-\bm g}^\parallel} & = \frac{a^2}{2L^2} (\lambda + 2 \mu) u_{\bm g}^\parallel - \frac{a\bm g}{iLg^2} \cdot \left( \frac{\partial V}{\partial \bm \phi} \right)_{\bm g}, \label{eq:hetero1} \\
    \frac{\partial f}{\partial u_{-\bm g}^\perp} & = \frac{a^2}{2L^2} \mu u_{\bm g}^\perp - \frac{a \hat z \times \bm g}{iLg^2} \cdot \left( \frac{\partial V}{\partial \bm \phi} \right)_{\bm g}. \label{eq:hetero2}
\end{align}
The energy is minimized for
\begin{align}
    \frac{a}{L} u_{\bm g}^\parallel & = \frac{2 \bm g}{(\lambda + 2 \mu)ig^2} \cdot \left( \frac{\partial V}{\partial \bm \phi} \right)_{\bm g}, \\
    \frac{a}{L} u_{\bm g}^\perp & = \frac{2 \hat z \times \bm g}{\mu ig^2} \cdot \left( \frac{\partial V}{\partial \bm \phi} \right)_{\bm g},
\end{align}
with $\bm g = M^\top \bm b$ and explicitly,
\begin{align}
    \left( \frac{\partial V}{\partial \bm \phi} \right)_{\bm g} & = i \sum_{\bm b'} \bm b' V_{\bm b'} \, \frac{1}{A_\mathrm{m}} \int_\text{moir\'e cell} d^2\bm r \, e^{i \bm b' \cdot \bm \phi(\bm r) - i \bm b \cdot M \bm r} \\
    & = i \sum_{\bm b'} \bm b' V_{\bm b'} \int_0^1 ds_1 \int_0^1 ds_2 \, e^{i (\bm b' - \bm b) \cdot (s_1 \bm a_1 + s_2 \bm a_2 ) + i \bm b' \cdot \bm u(s_1 \bm L_1 + s_2 \bm L_2)} \\
    & = i \sum_{\bm b'} \bm b' V_{\bm b'} \int_0^1 ds_1 \int_0^1 ds_2 \, e^{i (\bm b' - \bm b) \cdot (s_1 \bm a_1 + s_2 \bm a_2 ) + i \bm b' \cdot \sum_{\bm b''} \bm u_{\bm b''} e^{i\bm b'' \cdot (s_1 \bm a_1 + s_2 \bm a_2)}}.
\end{align}
\end{widetext}

\begin{figure*}
    \centering
    \includegraphics[width=.8\linewidth]{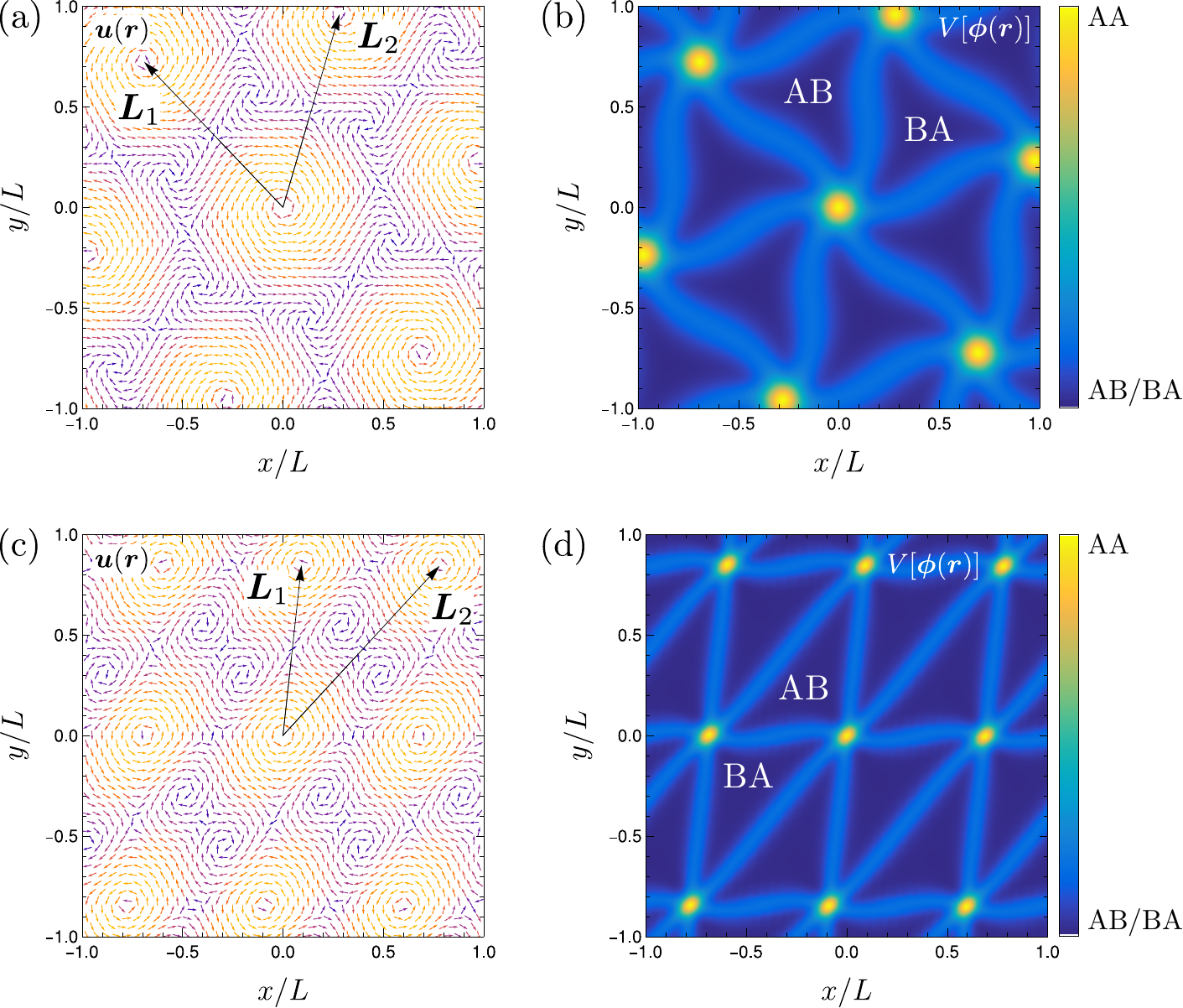}
    \caption{(a) Displacement field $\bm u(\bm r) = \bm u_1(\bm r) - \bm u_2(\bm r)$ of the relaxed configuration obtained by numerically solving the self-consistent equations \eqref{eq:hetero1} and \eqref{eq:hetero2} for $\theta = 0.3^\circ$ and 5\% biaxial heterostrain. (b) Local adhesion energy $V[\bm \phi(\bm r)]$ corresponding to (a). (c) Same as (a) for 3\% uniaxial heterostrain along the $x$ axis. (d) Local adhesion energy corresponding to (c). Here we used parameters for twisted bilayer graphene from Ref.\ \cite{carr_relaxation_2018}.}
    \label{fig:heterostrain}
\end{figure*}

We now consider two examples of heterostrain in twisted bilayer graphene for $\theta = 0.3^\circ$ using the parameters of Ref.\ \cite{carr_relaxation_2018} as an example. In Fig.\ \ref{fig:heterostrain}(a) and (b) we show the displacement field of the relaxed configuration as well as the local adhesion energy in the presence of biaxial heterostrain $\mathcal E_\text{biaxial} = \epsilon \sigma_0$ for $\epsilon = 0.05$. In this case, the moir\'e still preserves $C_6$ symmetry. We observe a swirly pattern due to the competition of two moir\'e scales \cite{mesple_giant_2023}. Finally, we consider uniaxial heterostrain \cite{kang_analytical_2025},
\begin{equation}
    \mathcal E_\text{uniaxial} = R(-\phi) \begin{pmatrix} \epsilon_1 & 0 \\ 0 & \epsilon_2 \end{pmatrix} R(\phi),
\end{equation}
with $\phi = 0$, $\epsilon_1 = 0.03$, and $\epsilon_2 = -\sigma \epsilon_1$ with $\sigma = 1 / \left( 1 + 2 \mu / \lambda \right) \approx 0.188$ the Poisson ratio of graphene \cite{carr_relaxation_2018}. The results are shown in Fig.\ \ref{fig:heterostrain}(c) and (d).

\section{\textsc{lammps} molecular dynamics simulations}

In the regime of marginal twist angles, the moir\'e lattice of a twisted 2H TMD homobilayer contains $6/\theta^2$ atoms, e.g., about $3 \times 10^5$ for $\theta = 0.25^\circ$. In this regime first-principle calculations are prohibitively expensive. Therefore we model atomic relaxation with molecular dynamics simulations using the Large-scale Atomic/Molecular Massively Parallel Simulator (\textsc{lammps}) code which employs classical interatomic force field models \cite{thompson_lammps_2022}. While these molecular dynamics simulations allow for larger supercell sizes, they have inherent limitations on accuracy over the choice of interatomic potentials. In our experience, while different interatomic potentials might give slightly different numerical values for the $c_i = V_i / \mu$ parameters, their qualitative behavior and symmetry properties are identical. 

For tWSe$_2$ we use the Kolmogorov–Crespi potential for interlayer interactions \cite{naik_kolmogorovcrespi_2019} and the Stillinger–Weber (SW) potential for intralayer interactions with SW/mod style \cite{jiang_parametrization_2015}. In this work, we performed relaxation calculations for commensurate twist angles ranging from $\theta = 0.1^\circ$ to $\theta = 10^\circ$. The smallest twist angles correspond to moir\'e cells with over one million atoms. Despite the large number of atoms in the simulation cell, the structural optimization remains computationally tractable due to the low cost of the classical potentials.

To compare with continuum elasticity which only describes acoustic degrees of freedom, we first extract the center-of-mass motion:
\begin{equation}
    \bm u_l = \sum_i \frac{m_i}{M} \bm u_{li},
\end{equation}
for each layer $l=1,2$. Here $M = \sum_i m_i$ and the sum runs over atoms in the monolayer unit cell. For TMDs MX$_2$, we obtain
\begin{equation}
    \bm u_l = \frac{m_\text{M}}{M} \bm u_{l,\text{M}} + \frac{m_\text{X}}{M} \left( \bm u_{l,\text{X1}} + \bm u_{l,\text{X2}} \right),
\end{equation}
with $M = m_\text{M} + 2 m_\text{X}$. For WSe$_2$ we use $m_\text{W} / M = 0.538$ and $m_\text{Se} / M = 0.231$ and for MoTe$_2$ with the data of Ref.\ \cite{zhang_polarization-driven_2024} we use $m_\text{Mo} / M = 0.273$ and $m_\text{Te} / M = 0.363$.

\begin{figure}
    \centering
    \includegraphics[width=\linewidth]{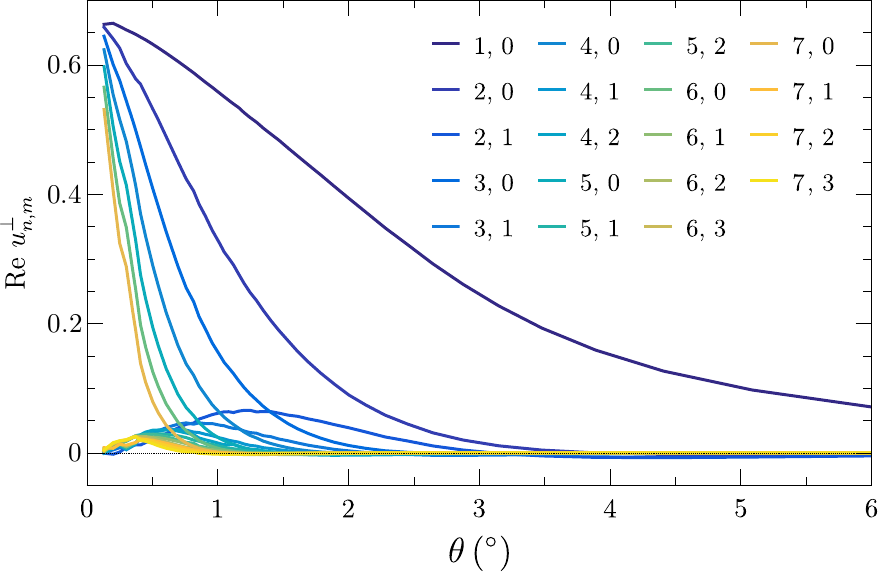}
    \caption{Real part of transverse Fourier components for tWSe$_2$ near $0^\circ$ (parallel stacking) calculated from \textsc{lammps} simulations. We do not show the imaginary part and the longitudinal components because they are over two orders of magnitude smaller. Only distinct values not related by symmetry are shown for the first $7$ shells ($28$ stars).}
    \label{fig:WSe20}
\end{figure}

\begin{figure}
    \centering
    \includegraphics[width=\linewidth]{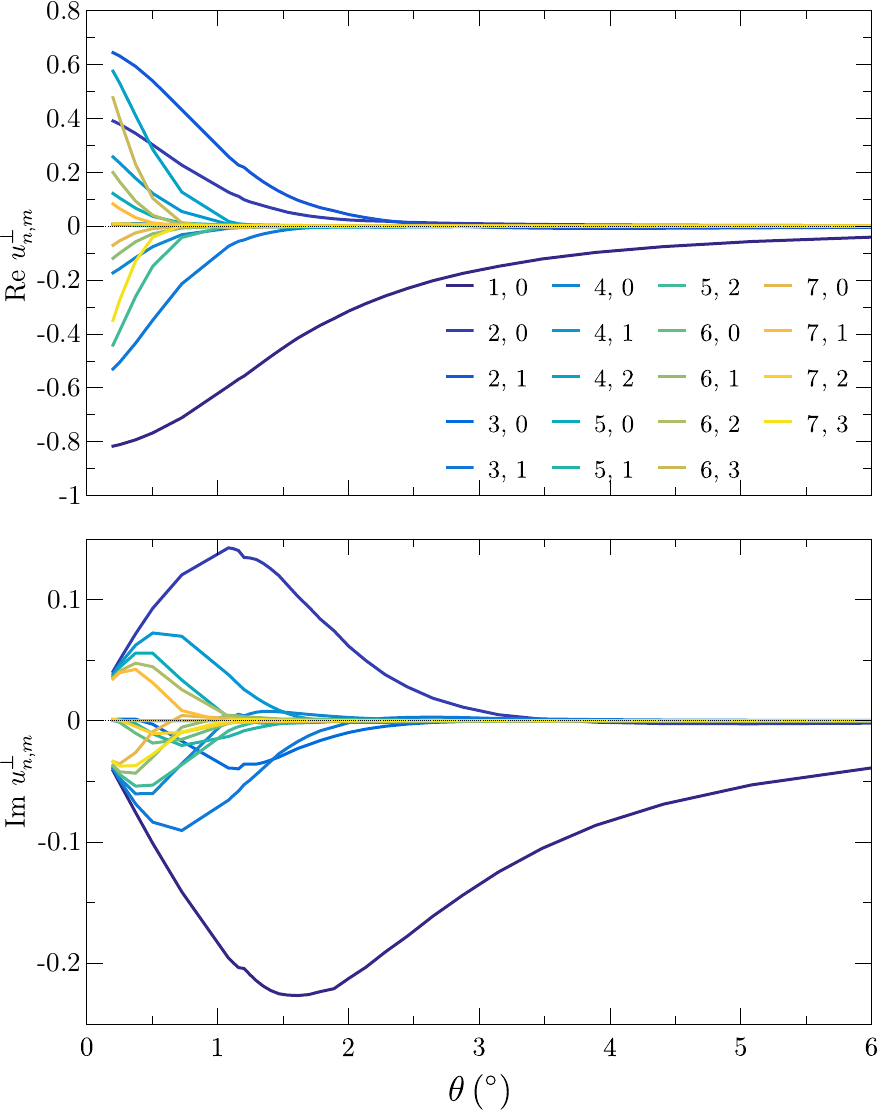}
    \caption{Transverse Fourier components for tWSe$_2$ near $60^\circ$ (antiparallel stacking) calculated from \textsc{lammps} simulations. Here the total twist angle is $60^\circ + \theta$. We do not show the longitudinal components because they are over two orders of magnitude smaller.  Only distinct values not related by symmetry are shown for the first $7$ shells ($28$ stars).}
    \label{fig:WSe260}
\end{figure}

In Figs.\ \ref{fig:WSe20} and \ref{fig:WSe260} we show the transverse Fourier components of the interlayer acoustic displacement field $\bm u = \bm u_1 - \bm u_2$ for twist angles near $0^\circ$ and $60^\circ$, respectively. For twist angles near $0^\circ$ (parallel stacking) the stacking symmetry is only approximately given by $D_6$ such that there is a small imaginary part. However, because the latter is over two orders of magnitude smaller, we do not show it. Similarly the longitudinal part is almost three orders of magnitude smaller. On the other hand, for twist angles near $60^\circ$ (antiparallel stacking) the stacking symmetry is given by $D_3$ and the imaginary transverse component is significant in general. Only for marginal twist angles, does the imaginary component vanish as the honeycomb soliton network recovers the $D_6$ symmetry.

\end{document}